\documentclass[12pt,a4paper,twoside,openright,prb]{article} 
\usepackage{graphics} 
\usepackage{graphicx} 
\usepackage{caption2}
\newcommand{\be}{\begin{equation}} 
\newcommand{\ee}{\end{equation}} 
\newcommand{\ba}{\begin{eqnarray}} 
\newcommand{\ea}{\end{eqnarray}}

\begin{document} 
\baselineskip 0.65cm

\title{\bf Effects of dissipation on disordered quantum spin models} 
\vskip 20pt 
 
\author{ L. F. Cugliandolo$^{1,2}$, D. R. Grempel$^3$,  
G. Lozano$^{4,5}$ and H. Lozza$^5$ \\  
{\footnotesize \it $^1$ 
Laboratoire de Physique Th{\'e}orique de l'{\'E}cole Normale 
Sup{\'e}rieure,} \\  
{\footnotesize \it 24 rue Lhomond, 75231 Paris 
Cedex 05, France} \\  
{\footnotesize \it $^2$ Laboratoire de Physique 
Th{\'e}orique et Hautes {\'E}nergies, Jussieu, } \\ {\footnotesize \it 
1er {\'e}tage, Tour 16, 4 Place Jussieu, 75252 Paris Cedex 05, France} 
\\  
{\footnotesize \it $^3$CEA-Saclay/SPCSI, 91191 Gif-sur-Yvette 
CEDEX, France }\\  
{\footnotesize \it $^{4}$ Department of Mathematics, 
Imperial College London,}\\ 
{\footnotesize\it 180 Queen's Gate, London SW7 2BZ, United Kingdom} 
\\ 
{\footnotesize \it 
$^5$ Departamento de F\'{\i}sica, FCEyN,  
Universidad de Buenos Aires, 
} 
\\ 
{\footnotesize \it 
Pabell{\'o}n I, Ciudad Universitaria, 1428 Buenos Aires, Argentina} 
} 
 
\date\today 
\maketitle 

\abstract{We study the effects of the coupling to an Ohmic quantum
reservoir on the static and dynamical properties of a family of
disordered $SU(2)$ spin models in a transverse magnetic field using a
method of direct spin summation. The tendency to form a glassy phase
increases with the strength of the coupling of the system to the
environment. We study the influence of the environment on the features
of the phase diagram of the various models as well as the stability of
the possible phases.}
 
\addtolength{\baselineskip}{0.2\baselineskip}

\section{Introduction} 
\label{sec:intro} 
The coupling of quantum two-level systems ({\sc tls}) to a dissipative
environment has decisive effects on their dynamical properties. The
case of dilute systems, in which interactions between the {\sc tls}
can be neglected, has been extensively investigated in the
literature~\cite{review-Leggett,Weiss} and is now well understood. The
physics that emerges when interactions between the {\sc tls} may not
be neglected has received much less attention.
   
In this paper we study the effects of a dissipative environment on 
the equilibrium and dynamical properties  of quantum {\em 
glassy} systems.  While spin-glass phases in which quantum 
fluctuations play an important role were found in a number of 
experimental systems~\cite{aeppli,lohn,tabata}, the 
 importance of interactions on the properties of tunneling defects 
in structural glasses was recently demonstrated~\cite{ladieu}.    
 
Mean-field~\cite{mean-field} and finite
dimensional~\cite{review-quantum-sg} quantum spin-glass models have
been studied in detail in the last few years.  The interest in the
effects of the coupling to the environment on the properties of
quantum glassy systems is more recent. The quantum spherical $p$-spin
glass model~\cite{Culo,Cugrlolosa,Chandra}, the $SU(N)$ random
Heisenberg model in the limit $N\to\infty$~\cite{Bipa}, and the
quantum random walk problem~\cite{Pottier} were studied in detail.  In
these cases the relevant degrees of freedom are continuous which makes
analytical treatments possible.  Here we analyze the more realistic
case of quantum $SU(2)$ spins with infinite range interactions between
$p$-uplets of spins coupled to a bath of harmonic oscillators.  For
$p=2$ this model reduces a model for metallic spin-glasses studied
previously~\cite{Kondo}. For $p > 3$ the model exhibits richer
behavior~\cite{Niri,Cugrsa1,Cugrsa2,Bicu}, including the possibility
of having first order transition lines.
 
We use a method of direct spin summation ({\sc dss}) first used in the
context of disordered spin models by Goldschmidt and Lai~\cite{Gold}.
In this method the disorder-averaged free-energy density is computed
using the replica method to  average over the random
quenched interactions. Next, a Trotter decomposition is performed in
order to express the partition function of the resulting single-site
self-consistent problem as a sum over different contributions, each
coming from a possible spin history $\sigma(\tau)=\pm 1$.  The
continuous imaginary-time variable $0 \le \tau \le \beta\hbar $ is
discretized on a grid $\tau_t = t \beta\hbar /M$, $t=0,\cdots,M-1$,
and the partition function is computed by numerically performing the
{\it exact} sum over the $2^M$ possible discrete spin histories,
$\sigma_t \equiv \sigma(\tau_t)=\pm 1$. Physical results are obtained
repeating this procedure for various values of $M$ and extrapolating
to $M\to\infty$.

 We find that the coupling to the enviromement favors the appearance of
the spin-glass phase reducing the strength of the quantum fluctuations
that tend to destabilize it.  For $p=2$ the phase transition is always
second order. For $p \ge 3$ there exists a tricritial temperature
$T^\star$ below which quantum fluctuations drive the transition first
order. $T^\star$ decreases with the strength of the coupling to the
bath.  For $p \ge 3$ a dynamic transition precedes the equilibrium phase
transition. The coupling to the bath also stabilizes the dynamic
glassy phase.
 
The organization of the paper is as follows. In Sect.~\ref{sec:model}
we introduce the coupled $p$-spin-bath model and the formalism that we
use to solve it.  In Sect.~\ref{sec:results} we discuss the numerical
method and present our results. Sect.~\ref{sec:conc} contains our
conclusions.

\section{The model} 
\label{sec:model} 
 
We are interested in disordered $SU(2)$ spin models in a 
transverse field described by Hamiltonians of the type 
\begin{equation} 
H_s=H_L - \Gamma \sum_{i=1}^{N} \hat{\sigma}_i^x 
\end{equation} 
with 
\begin{equation} 
H_L= -\sum_{i_1 < \cdots<i_p}^{N} 
J_{i_1, \dots, i_p} \hat{\sigma}_{i_1}^z \cdots \hat{\sigma}_{i_p}^z 
\; . 
\end{equation} 
Here, $\hat \sigma^x,\hat \sigma^y,\hat \sigma^z$ are the standard 
Pauli matrices, $J_{i_1 \cdots i_p}$ denotes a quenched random exchange 
between $p$ spins, $\Gamma$ is the transverse field introducing 
quantum fluctuations and $N$ is the total number of spins.  The sum 
runs over all possible $p$-uplets of spins. The model is then 
fully-connected and mean-field in character.  It is completely 
determined once the order $p$ and the precise distribution 
$\mathcal{P}[J] $ of random interactions are chosen. We consider the 
case in which the random independent variables $J_{i_1,\dots ,i_p}$ 
are Gaussian with zero mean and variance $p!{J}^2/2 N^{p-1}$. The 
scaling of the variance with $N$ is chosen so as to ensure a good 
thermodynamic limit. 
 
We study  the thermodynamics and some aspects of the non-equilibrium 
dynamics of the quantum spin model coupled to a quantum environment 
 assumed to be in thermal equilibrium. 
We model the coupling to the environment by assuming that each spin in the 
system is coupled 
to its own set of $\tilde{N}/N$ independent harmonic oscillators with  
$\tilde N$ the total number of them. 
The bosonic Hamiltonian for the isolated reservoir is 
\begin{equation} 
H_b=\sum_{l=1}^{\tilde{N}} \frac1{2m_l} \hat{p}_{l}^2 + 
\sum_{l=1}^{\tilde{N}} \frac12 m_{l} \omega_{l}^2 \hat{x}_{l}^2 
\; . 
\end{equation} 
The coordinates $\hat x_l$ and the momenta $\hat p_l$ 
 satisfy canonical 
commutation relations. 
For simplicity we consider a bilinear coupling, 
\begin{equation} 
H_{sb} = -\sum_{i=1}^N  \hat{\sigma}_i^{z} \sum_{l=1}^{\tilde{N}} 
c_{i l} \hat{x}_{l} 
\; , 
\end{equation} 
that involves only the oscillator coordinates. 
 The Hamiltonian for the coupled system is then given by 
\begin{equation} 
H=H_s+H_b+H_{sb}+H_{ct} 
\end{equation} 
where we added a counter-term, 
\begin{equation} 
H_{ct}=\sum_{l=1}^{\tilde{N}}\frac{1}{2 m_l \omega_{l}^2} 
\left(\sum_{i=1}^{N} c_{i l} \hat{\sigma}_i^z \right)^2 
\; , 
\label{eqmodels} 
\end{equation} 
whose effect is to eliminate a  possible mass-normalization 
induced by the coupling to the bath~\cite{Weiss}. 
 
The partition function of the combined system 
 for a particular realization of 
the bonds  
\begin{equation} 
Z=\mbox{Tr}[e^{- \beta H}] 
\end{equation} 
involves a sum over all states of the system and of the bath.  The
trace over the variables of the bath can be performed explicitly using
standard techniques~\cite{Weiss,Feve,Leggett}.  The result of tracing
out these variables can be expressed in terms of the spectral function
of the bath,
\begin{eqnarray} 
I_{ij}(\omega) &\equiv& 
\frac{\pi}{2}\sum_{l=1}^{\tilde{N}}\frac{c_{il} c_{jl}}{m_l \omega_{l}} 
\delta(\omega- \omega_{l}) 
= 
\delta_{ij} I(\omega) 
\; . 
\end{eqnarray} 
We chose to study the effect of an Ohmic bath 
parameterized as 
\begin{equation} 
I(\omega)=2 \eta \hbar \omega \Theta(\omega_{max}-\omega) 
\end{equation} 
with $\eta$ the friction constant, $\omega_{max}$  an ultra-violet cut-off 
and $\Theta(x)$ the Heaviside theta-function. 
 
This problem can be mapped onto a classical Ising spin system using
the Totter-Suzuki formalism~\cite{Niri,Gold,Grempel}.  This amounts to
writing the path-integral for the partition function as a sum over
spin and oscillator variables evaluated on a discrete imaginary-time
grid, $\tau_t=\beta\hbar/M t$, labeled by the index $t=0,\dots,M-1$.
Periodic boundary conditions on the discrete time-axis are imposed due
to the fact that the evaluation of the partition function involves a
trace.  To recover the correct representation of the trace the limit
$M\to\infty$ should be ultimately taken. The finite $M$ expression
yields a sequence of $M$-approximants to the asymptotic $M\to\infty$
formula.
 
The $M$-th approximant of the ``reduced'' partition function obtained
after integrating out the bath reads
\begin{eqnarray} 
Z&=& 
\mbox{Tr}_{\{\sigma_i^t\}} \exp \left[\frac{\beta}{M} 
\sum_{t=0}^{M-1} 
\sum_{i_1 <\cdots <i_p}^{N} J_{i_1 \cdots i_p} \sigma_{i_1}^{t} 
\cdots \sigma_{i_p}^{t}  + 
\sum_{t=0}^{M-1} \sum_{i=1}^{N} \left( A+B\sigma_i^{t}\sigma_i^{(t+1)} 
\right) \right. 
\nonumber \\ 
& & \mbox{\hspace{2cm}}  \left. - \sum_{t,t'=0}^{M-1} \sum_{i=1}^{N}  
(1-\sigma_i^{t} \sigma_i^{t'}) 
 \mathcal{C}_{(t-t')} \right] 
\end{eqnarray} 
where 
\begin{eqnarray} 
A &=& \frac{1}{2} \ln \left[ \sinh \left(\frac{\beta \Gamma}{M}  \right)  
\cosh \left(\frac{\beta \Gamma}{M}  \right) \right] 
\, , 
\\ 
B &=& \frac{1}{2} \ln \left[ \coth \left(\frac{\beta \Gamma}{M}  \right) 
\right] 
\; , 
\\ 
\mathcal{C}_{(t-t')} &=& 
\frac{2\eta}{\pi \hbar} \int_0^{\omega_{max}} \! d\omega \; \frac{ \cosh 
\left(\omega \beta \hbar \left( (t-t')/M-1/2 \right) \right) 
\sinh^2 \left(\omega 
\beta \hbar /2M \right)}{\omega \sinh\left(\omega \beta \hbar / 2 \right)} 
\; . 
\end{eqnarray} 
The trace represents the sum over all $2^{N \times M}$ distinct
classical Ising spin configurations, $\sigma_i^t=\pm 1$, for each
spin, $i=1,\dots, N$, evaluated at each time-slice, $t=0,\dots,M-1$.
 
The disordered averaged free-energy is calculated using 
the replica trick~\cite{Mepavi} 
\begin{equation} 
\beta \overline F 
= - \overline{\ln Z } = - \lim_{n \rightarrow 0} \frac{\overline{Z^n} -1}{n} 
\; . 
\label{eqreplica} 
\end{equation} 
After some standard manipulations, and up to some irrelevant factors, 
we obtain 
\begin{equation} 
\overline{Z^n} = 
\prod_{a,b=1}^{n} \prod_{t,t'=0}^{M-1} \int \mathcal{D}Q^{atbt'} 
\mathcal{D}\Lambda^{atbt'}  
\exp \left( -N P(\Lambda, Q) \right) 
\end{equation} 
with 
\begin{eqnarray} 
&& 
P(\Lambda, Q)= 
\sum_{a,b=1}^{n} \sum_{t,t'=0}^{M-1} 
\left(   \frac{\Lambda^{atbt'}}{M^2} Q^{atbt'} - 
\frac{\beta^2 J^2}{4M^2} (Q^{atbt'})^{\bullet p} + 
\mathcal{C}_{(t-t')}\delta^{ab}(1- Q^{atbt'})\right) 
\nonumber \\ 
&& \;\;\;\;\;\;\;\;\;\;\;\;\;\;\;\;\;\; 
- \ln \mbox{Tr}_{\{\sigma^{at}\}} e^{H_{ eff}} 
\; , 
\label{eqP} 
\\ 
&& 
H_{ eff} = 
\sum_{a,b=1}^{n} \sum_{t,t'=0}^{M-1} 
\frac{\Lambda^{atbt'}}{M^2} \sigma^{at} \sigma^{bt'} + 
\sum_{a=1}^{n} \sum_{t=0}^{M-1} 
\left(A +  B \sigma^{at} \sigma^{a(t+1)} \right) 
\; . 
\label{eqF} 
\end{eqnarray} 
where the bullet is used to distinguish the ordinary power from the
matrix power.  In the thermodynamic limit, $N\to\infty$, the integrals
in $\overline Z^n$ can be evaluated with the saddle point method at
the expense of exchanging the $N\to\infty$ and $n\to 0$ limits. The
disordered-averaged free-energy per spin is then
\begin{equation} 
\beta \overline f = - \lim_{n \rightarrow 0} \frac{P[\Lambda_0,Q_0]}{n} 
\label{eqf} 
\end{equation} 
where 
$\Lambda_0$ and $Q_0$ are such that, 
\begin{equation} 
\left.\frac{\delta P(Q,\Lambda)}{\delta Q} \right|_{Q_0,\Lambda_0}=0 \; , 
\,\,\,\,\,\,\,\,\,\,\, 
\left.\frac{\delta P(Q,\Lambda)}{\delta \Lambda} \right|_{Q_0,\Lambda_0}=0 
\; . 
\end{equation} 
Hereafter we omit subscripts in the saddle-point values $Q_0$ and 
$\Lambda_0$.  
The disorder-averaged entropy per spin is easily obtained from the 
disorder-averaged free-energy density 
and reads 
\begin{eqnarray} 
-\frac{\overline s}{k_B}  &=& \beta \overline f + 
\frac{1}{n} \sum_{a,b=1}^n \sum_{t,t'=0}^{M-1} \left[ 
\frac{\beta^2 J^2}{2 M^2} 
 (Q^{atbt'})^{\bullet p} 
- 
\delta^{ab}(1- Q^{atbt'}) 
\frac{\partial \mathcal{C}_{(t-t')}}{\partial \beta} 
\right] 
\nonumber \\ 
& & 
+ \frac{\beta \Gamma}{\sinh(2\beta \Gamma /M)} 
\left[\cosh(2\beta \Gamma /M)-\frac{1}{n} \sum_{a=1}^n Q^{a(t+1)at}\right] 
\; . 
\label{eq:entropy} 
\end{eqnarray} 
Another physical observable of interest is 
the magnetic susceptibility, 
\begin{equation} 
\chi = \left. \frac{\partial \mathcal{M}}{\partial h}\right|_{h=0} 
\end{equation} 
where $\mathcal{M}=N^{-1} \sum_{i=1}^N \overline{\langle
\sigma_i^z\rangle} $ is the total disorder-averaged magnetization and
$h$ a longitudinal external magnetic field. In terms of $Q^{atbt'}$
the susceptibility is given by
\begin{equation} 
\chi = \frac{\beta}{M^2} \sum_{a,b=1}^n \sum_{t,t'=0}^{M-1} Q^{atbt'} 
\; . 
\label{eq:susc} 
\end{equation} 
In Eqs.~(\ref{eq:entropy}) and (\ref{eq:susc}) 
the right-hand-sides should be evaluated at the saddle-point 
values. 
 
The matrix elements $Q^{atbt'}$ are the order parameters of the 
 model, 
\begin{equation} 
Q^{atbt'} = 
\frac1{N} \sum_{i=1} \overline{\langle \sigma_i^{at} \sigma_i^{bt'}} 
\rangle 
\; . 
\end{equation} 
Because of the translational invariance in the Trotter time 
 direction the diagonal terms in the replica indices 
depend on the time-difference only, 
\begin{equation} 
Q^{atat'}=q_d(t-t') 
\;. 
\end{equation} 
Notice that due to the periodic boundary condition $q_d(t)=q_d(M-t)$. 
In addition, as $q_d(0)=1$, only 
$(t-t')=1,2,\cdots,\mbox{int}\frac{M}{2}$ need to be considered.  The 
off-diagonal elements in the replica indices, $Q^{atbt'}$ with $a\neq b$, 
are $t$ and $t'$ independent as shown by Bray 
and Moore~\cite{Bray-Moore}. 
 
In order to determine the different phases of the model, we consider 
the following {\it Ans\"atze}: 
 
\vspace{.25cm} 
\noindent 
\underline{\it Paramagnetic Phase} 
\vspace{.25cm} 
 
The matrices $Q$ and  $\Lambda$ are  taken to be  diagonal in replica space 
\begin{equation} 
Q^{atbt'} = q_d (t-t') \delta^{ab} \; , 
\;\;\;\;\;\;\;\;\; 
\Lambda^{atbt'} = \lambda_d (t-t') \delta^{ab} 
\; . 
\end{equation} 
Using Eqs.~(\ref{eqP}), 
(\ref{eqF}) and (\ref{eqf}) the disorder-averaged 
free-energy per spin can be expressed as 
\begin{equation} 
\beta \overline f= 
\frac{\beta^2 J^2}{4 M^2} (p-1) 
\sum_{t\not=t'}^{M-1} 
 q_d^p (t-t') - 
\frac{\beta^2 J^2}{4 M} + 
\sum_{t \not =t'}^{M-1} \mathcal{C}_{(t-t')}  - 
\ln\mbox{Tr}_{\{\sigma^t\}} e^{H^{ pm}_{ eff}} 
\end{equation} 
with 
\begin{equation} 
H^{ pm}_{ eff} 
= 
\frac{1}{M^2} 
\sum_{t \not = t'}^{M-1} \lambda_d (t-t') \sigma^t \sigma^{t'} + 
\sum_{t=0}^{M-1}(A+B \sigma^t \sigma^{(t+1)}) 
\; . 
\end{equation} 
Here, 
$ q_d (t-t')$ and $\lambda_d (t-t')$ are obtained self-consistently 
from the extremum condition  by summing over 
all $2^M$ spin configurations $\sigma^t=\pm 1$: 
\begin{eqnarray} 
q_d (t-t') &=& 
\frac{\mbox{Tr}_{\{\sigma^t\}} \left[e^{H^{ pm}_{ eff}} \sigma^t 
\sigma^{t'} \right]}{\mbox{Tr}_{\{\sigma^t\}} e^{H^{ pm}_{ eff}}} 
\; , 
\\ 
\lambda_d (t-t') &=& 
\frac{\beta^2 J^2 p}{4} q_d^{p-1} (t-t') + M^2 
 \mathcal{C}_{(t-t')} 
\; . 
\end{eqnarray}

\vspace{.25cm} 
\noindent 
\underline{\it Equilibrium spin-glass phase} 
\vspace{.25cm} 
 
In order to characterize this phase we use a one-step replica symmetry
breaking ({\sc rsb}) {\it Ansatz},
\begin{eqnarray} 
Q^{atbt'} &=& 
(q_d (t-t') -q_{ea})\delta^{ab} + q_{ea} \epsilon^{ab} 
\; , 
\\ 
\Lambda^{atbt'} &=& (\lambda_d (t-t') - \lambda_{ea}) \delta^{ab}+ 
\lambda_{ea} \epsilon^{ab} 
\; . 
\end{eqnarray} 
Using  Eqs.~(\ref{eqP}), (\ref{eqF}) 
and  (\ref{eqf}) the disordered-averaged 
free-energy density becomes 
\begin{eqnarray} 
\beta \overline f & = & \frac{\beta^2 J^2}{4 M^2}(p-1) 
\sum_{t\not=t'}^{M-1}   
q_d^p (t-t') + 
(m-1)\frac{\beta^2 J^2}{4}(p-1) q_{ea}^p - 
\frac{\beta^2 J^2}{4 M} \nonumber \\ 
& &   - \frac{1}{m}\ln \int  
d\mathcal{X} \; \left(\mbox{Tr}_{\{\sigma^t\}} 
e^{H^{ esg}_{ eff}}\right)^m 
+ 
\sum_{t \not =t'}^{M-1} \mathcal{C}_{(t-t')} 
\end{eqnarray} 
with 
\begin{eqnarray} 
H^{ esg}_{ eff} 
&=& 
\frac{1}{M^2} 
\sum_{t\not=t'}^{M-1} (\lambda_d (t-t')-\lambda_{ea}) 
\sigma^t \sigma^{t'}-\frac{\lambda_{ea}}{M} + 
 \frac{\sqrt{2 \lambda_{ea}}}{M} x \sum_t^{M-1} \sigma^t  
\nonumber\\ 
&& + 
\sum_t^{M-1}\left(A+B\sigma^t \sigma^{(t+1)}\right) 
\label{Heff} 
\end{eqnarray} 
and the integration measure 
\begin{equation} 
d{\mathcal X} \equiv 
\frac{dx}{\sqrt{2\pi}} e^{-\frac{x^2}{2}} 
\; . 
\end{equation} 
Here and in what follows all integrals over $x$ go from $-\infty$ 
to $\infty$. 
 
As in the paramagnetic phase, the order parameters $ q_d (t-t')$, 
$q_{ea}$, $\lambda_d (t-t')$ and $\lambda_{ea}$ are determined 
self-consistently from the extremum conditions that involve a 
sum over all $2^M$ 
spin configurations, $\sigma^t=\pm 1$: 
\begin{eqnarray} 
q_d(t-t')&=& 
\frac{\int  
d{\mathcal X} \; 
\left( \mbox{Tr}_{\{\sigma^t\}} 
e^{H^{ esg}_{ eff}} \right)^{m-1} 
\left(\mbox{Tr}_{\{\sigma^t\}} 
e^{H^{ esg}_{ eff}} \sigma^t \sigma^{t'} \right)} 
{\int  
d{\mathcal X} \; 
\left( \mbox{Tr}_{\{\sigma^t\}} e^{H^{ esg}_{ eff}} \right)^m} 
\; , 
\label{eq:qd} 
\\ 
q_{ea} &=& 
\frac{\int 
d{\mathcal X} \; 
\left( \mbox{Tr}_{\{\sigma^t\}} e^{H^{ esg}_{ eff}} \right)^{m-2} 
\left(\mbox{Tr}_{\{\sigma^t\}} 
e^{H^{ esg}_{ eff}} 
\sum_t \sigma^t / M\right)^2}{\int 
d{\mathcal X} \; \left( \mbox{Tr}_{\{\sigma^t\}} e^{H^{ esg}_{ eff}} 
\right)^m} 
\; , 
\label{eq:qea} 
\\ 
\lambda_d(t-t') &=& 
\frac{\beta^2 J^2 p}{4} 
q_d^{p-1}(t-t') 
+ M^2 \mathcal{C}_{(t-t')} 
\; , 
\\ 
\lambda_{ea}&=&\frac{\beta^2 J^2 p}{4} q_{ea}^{p-1} 
\label{eq:lambdaea} 
\; . 
\end{eqnarray} 
As it has been discussed in a number of papers on classical~\cite{pspins} 
and quantum~\cite{Niri,Cugrsa1,Cugrsa2,quantum-marg} 
spin-glass models, two choices for the determination of the 
breaking-point $m$ lead to different physical results. The use of the 
extremum condition, that corresponds to looking for the value of $m$ that 
renders the disorder-averaged 
free-energy stationary, leads to 
\begin{equation} 
m=I^{-1}\left[m^2 \frac{\beta^2 J^2 p}{4}(p-1) q_{ea}^p + 
\ln \int  
d{\mathcal X} \; 
\left(\mbox{Tr}_{\{\sigma^t\}} 
e^{H^{ esg}_{ eff}}\right)^m \right] 
\end{equation} 
where 
\begin{equation} 
I= \frac{\int  
d{\mathcal X} \; 
\left( \mbox{Tr}_{\{\sigma^t\}} 
e^{H^{ esg}_{ eff}} \right)^{m} 
\ln \left(\mbox{Tr}_{\{\sigma^t\}} 
e^{H^{ esg}_{ eff}}\right)}{\int   
d{\mathcal X} \; 
\left( \mbox{Tr}_{\{\sigma^t\}} 
e^{H^{ esg}_{ eff}} \right)^m} 
\; . 
\end{equation} 
With this choice one describes the equilibrium properties of the model. 
 
This {\it Ansatz} yields the exact solution~\cite{Cugrsa2} to the
spherical version of the $p\geq 3$ model.  When $p=2$ the spherical
model is solved by a simpler replica symmetric ({\sc rs}) form.  The
stability of the one-step {\it Ansatz} for quantum $SU(2)$ models can be
tested by extending the analysis of de Almeida and
Thouless~\cite{AT}. When the lowest eigenvalue, that is also called
the replicon, vanishes, the one-step {\sc rsb} {\it Ansatz} becomes
marginally stable. When the replicon is negative, this {\it Ansatz} is
unstable.
 
We do not expect the one-step {\sc rsb} {\it Ansatz} to be stable
everywhere in the phase diagram in the case of discrete spins.
Indeed, by evaluating the replicon at the order parameters and
break-point obtained from the extremum conditions we found that the
one-step {\sc rsb} {\it Ansatz} is unstable in the full spin-glass
phase when $p=2$ [Sherrington-Kirkpatrick ({\sc sk})] model indicating
the need to break the replica symmetry further.  In the case of the
$p\geq 3$ {\it }classical spin model, the one-step {\sc rsb} {\it
Ansatz} is unstable below a temperature $T_g < T_s$ as shown by
Gardner~\cite{Gardner} in the classical case. Thus, the solution for
the classical Ising $p$ spin model also requires full {\sc rsb} at
very low temperatures. $T_g$ depends on the parameter $p$ and, as
expected, it tends to $T_s=J$ when $p\to 2^+$ and it vanishes when
$p\to\infty$.
 
When quantum fluctuations are taken into account we thus expect to
find a Gardner line of instability.  A careful study of how this line
depends on $\Gamma$ and the coupling to the bath requires to solve the
quantum problem at rather low temperatures.  This is done in
Sect.~\ref{sec:results} where we compute the location of the Gardner
instability line. As seen in Fig.~\ref{fig:Gardner}, the region where
the one-step {\sc rsb} static {\it Ansatz} is unstable is quite small.
Outside this region the one-step {\sc rsb} {\it Ansatz} is exact and
can be studied to study the properties of the $p\geq 3$ quantum $SU(2)$
model. Elsewhere, and for $p=2$, we shall regard this solution as a
suitable approximation to the correct solution.
 
\vspace{.25cm} 
\noindent 
\underline{\it Dynamic spin-glass phase} 
\vspace{.25cm} 
 
The {\it marginality condition} leads to a different equation for
$m$. With this condition one requires that the saddle-point is only
marginally stable, {\it i.e.} the matrix of quadratic fluctuations has a zero
replicon eigenvalue (and one does not impose the condition of extreme
on $m$). It has been checked by comparison to the real-time
dynamics~\cite{Culo}, that this condition yields the freezing
transition of the spherical quantum $p$-spin model with $p\geq 3$
coupled to the oscillator reservoir at the initial time
$t=0$~\cite{Cugrlolosa}.  Here we use it as an indication of where
such a dynamic transition line should be located for the $SU(2)$ quantum
spin systems.
 
Adapting the calculation of de Almeida and Thouless~\cite{AT} 
to the quantum problem under study we find that the replicon eigenvalue 
is given by 
\begin{equation} 
\lambda_R=P-2Q+R 
\end{equation} 
with 
\begin{eqnarray} 
P&=&1-k q_{ea}^{p-2}t\; , 
\nonumber \\ 
Q&=&-k q_{ea}^{p-2}u 
\; , 
\nonumber \\ 
R&=&-k q_{ea}^{p-2}r 
\; . 
\end{eqnarray} 
The factors $r,u$ and $t$ are 
\begin{eqnarray} 
&& r = 
\langle \sigma^a \sigma^b \sigma^c \sigma^d \rangle  = 
\frac{\int d\mathcal{X} 
\left(\mbox{Tr}_{\sigma^t} e^{H_{ eff}^{ dsg}}\right)^{m-4} 
\left(\mbox{Tr}_{\sigma^t} e^{H_{ eff}^{ dsg}} 
(\sum_t \sigma^t/M)\right)^{4}} 
{\int d\mathcal{X} \left(\mbox{Tr}_{\sigma^t} 
e^{H_{ eff}^{ dsg}}\right)^{m}} 
\; , 
\\ 
&& u = 
\frac{1}{M^2} \sum_{t\tau} 
\langle \sigma^{at} \sigma^b \sigma^{a\tau} \sigma^d \rangle 
\\ 
&& \;\; = 
\frac{\int d\mathcal{X} 
\left(\mbox{Tr}_{\sigma^t} e^{H_{ eff}^{ dsg}}\right)^{m-3} 
\left(\mbox{Tr}_{\sigma^t} e^{H_{ eff}^{ dsg}} (\sum_t \sigma^t/M)\right)^{2} 
\mbox{Tr}_{\sigma^t} e^{H_{ eff}^{ dsg}} (\sum_{tt'} \sigma^t 
\sigma^{t'}/M^2)} 
{\int d\mathcal{X} \left(\mbox{Tr}_{\sigma^t} e^{H_{ eff}^{ dsg}}\right)^{m}} 
\nonumber 
\; , 
\\ 
&& t = 
\frac{1}{M^4} 
\sum_{tt'\tau \tau'} 
\langle \sigma^{at} \sigma^{bt'} \sigma^{a\tau} \sigma^{b\tau'} 
\rangle  
\nonumber \\ 
&& \;\; = 
\frac{\int d\mathcal{X} \left(\mbox{Tr}_{\sigma^t} 
e^{H_{ eff}^{ dsg}}\right)^{m-2} 
\left(\mbox{Tr}_{\sigma^t} 
e^{H_{ eff}^{ dsg}} (\sum_{tt'} \sigma^t \sigma^{t'}/M^2)\right)^2} 
{\int d\mathcal{X} \left(\mbox{Tr}_{\sigma^t} e^{H_{ eff}^{ dsg}}\right)^{m} 
} 
\; . 
\end{eqnarray} 
Here, we have defined  
\begin{equation} 
k\equiv\frac{\beta^2 J^2}{2}p(p-1) 
\; . 
\end{equation} 
and $H_{ eff}^{ dsg}$ is the Hamiltonian of by Eq.~(\ref{Heff}).
Finally, the values of $q_d(t-t')$ and $q_{ ea}$ are fixed by the
extremal conditions.
 
\section{Results} 
\label{sec:results} 
 
In this Section we describe the outcome of solving the equations we
derived in the previous Section and we discuss how the coupling to the
Ohmic bath of harmonic oscillators modifies the behavior of the spin
model.
 
\subsection{Numerical method} 
 
The free-energy density and derived magnitudes depend on the parameter
$M$ that in practice takes finite values.  Several strategies were
proposed to study the limit $M\to\infty$.  Usadel and
Schmitz~\cite{ussc} noted that $M$ should be such that $\beta \Gamma/M
\ll 1$. For low temperatures this criterium becomes quickly
impractical since one cannot perform the complete sum over states for
such large values of $M$. As an alternative, these authors proposed to
use a Montecarlo procedure to estimate the sum over configurations
when $M$ is large~\cite{ussc,alri}.
 
In this paper we use another method that has been previously used to
study the isolated quantum {\sc sk}~\cite{Gold} and the $SU(2)$
$p$-spin~\cite{Niri} models in a transverse field. We compute each
magnitude using a direct spin sumation ({\sc dss}) with $M$ ranging
form $M=8$ to $M=13$. The six values obtained are the input for a
polynomial extrapolation which gives the $M \rightarrow \infty $ limit
for the considered magnitud. For almost all cases, we extrapolate
using the $(1/M)^2$ law~\cite{suzuki}.  Special care is taken for the
study of the function $q_d(\tau)$ that is known on the imaginary-time
grid $0, \frac{\beta\hbar}{M}, \frac{2\beta\hbar}{M}, \cdots,
\beta\hbar$. We interpolate these sets of points with splines for each
value of $M$.  As the value of $M$ increases the density of points
also increases and we have a better description of the curve. Finally
we use the result of the function $q_d^M(\tau)$ at $\tau$ as the input
for the polynomial extrapolation. Because of the additional
difficulties introduced by the functional nature of $q_d(\tau)$ we
considered a linear extrapolating scheme in this case.
 
\begin{figure} 
\begin{center} 
\resizebox{11cm}{!} 
{\includegraphics*[1.5in,4.7in][7in,7.7in]{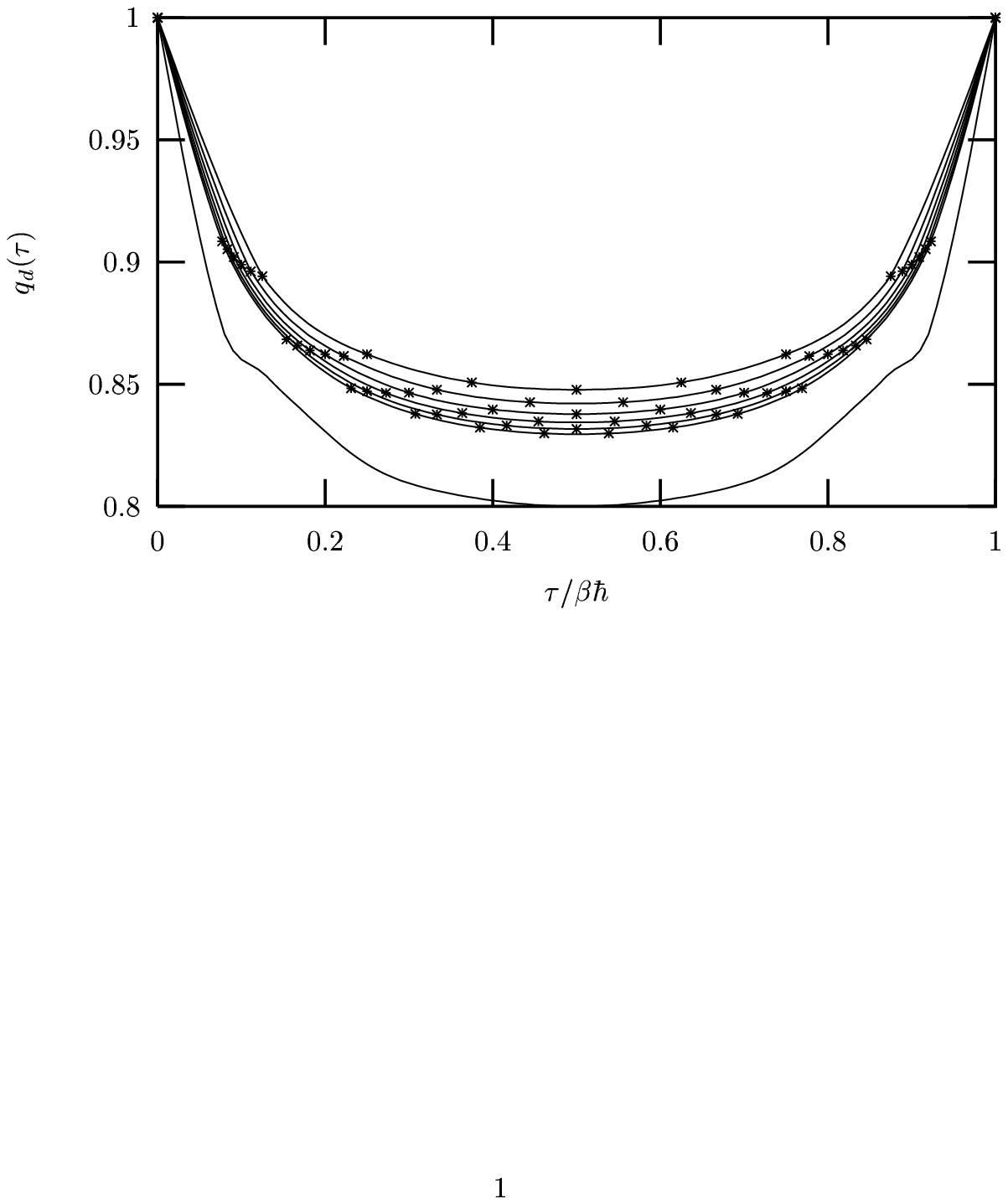}} 
\resizebox{11cm}{!} 
{\includegraphics*[1.5in,4.7in][7in,7.7in]{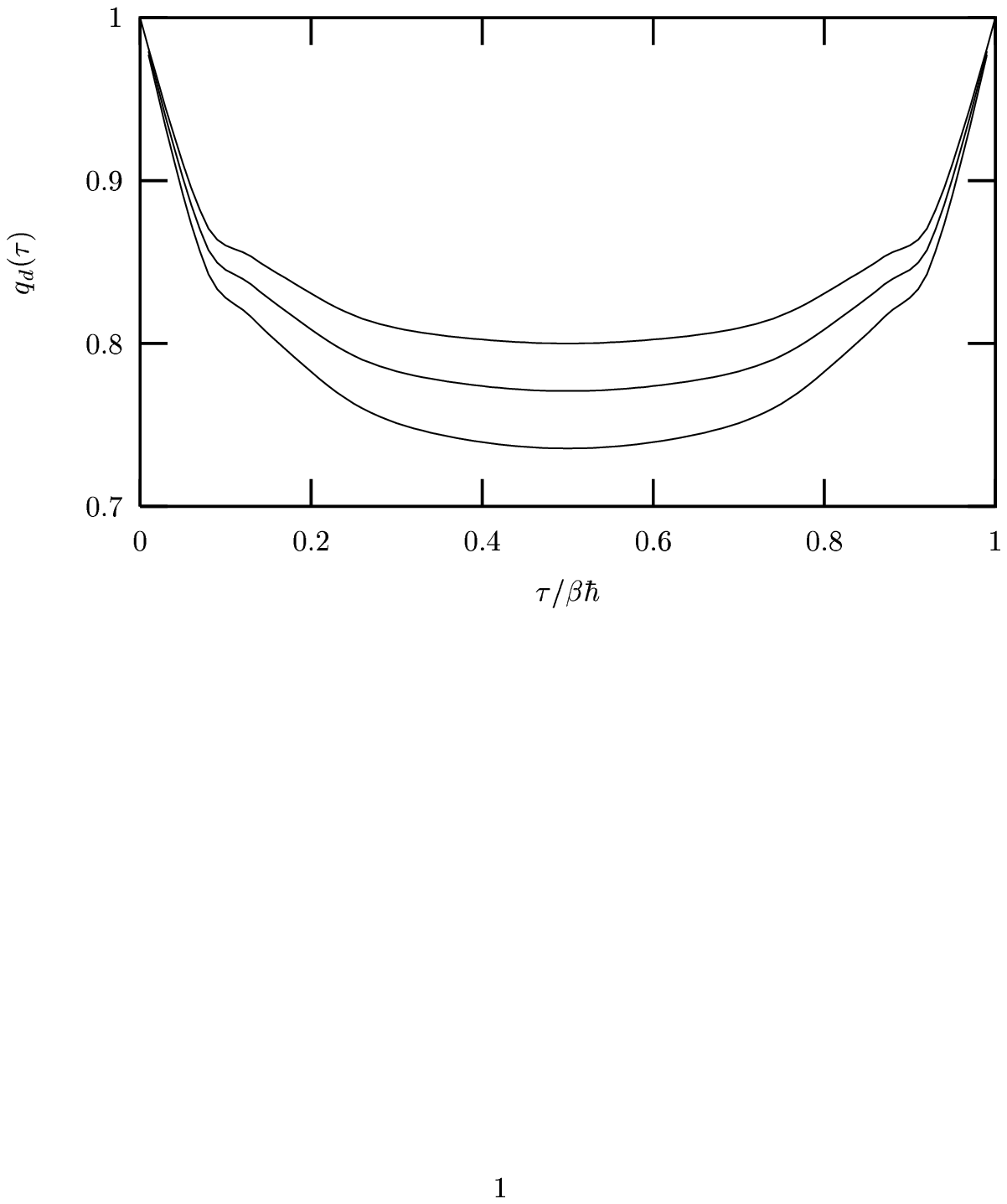}} 
\setcaptionwidth{13cm}
\caption{\footnotesize Upper panel: The diagonal imaginary-time
dependent order parameter $q_d(\tau)$ in the $p=3$ model as a function
of $\tau/(\beta\hbar)=t/M$ with $t$ the Trotter time,
$t=0,1,\cdots,M$. The temperature is $T =0.3$ and the transverse
magnetic field is $\Gamma = 0.8$. The coupling to the bath is
$\eta=1$. The solutions for $M=8,9,10,11,12,13$ are shown from top to
bottom, (with linepoints, the points being the numerical data and the
lines representing the result of splines) and the lowest curve (thin
line) is the result of the extrapolation to $M\to\infty$.  Lower
panel: The limiting curve $\lim_{M\to\infty} q^M_d(\tau)$ for three
couplings to the environment: $\eta=0$ (bottom), $\eta=0.5$ (middle)
and $\eta=1$ (top). The other parameters are the same as in the upper
panel.  (See the text for the details of the extrapolated method
used.)}
\label{fig:M-dep0} 
\end{center} 
\end{figure} 
 
\begin{figure} 
\begin{center} 
\resizebox{11cm}{!}{\includegraphics*[1.5in,4.7in][7in,7.7in] 
{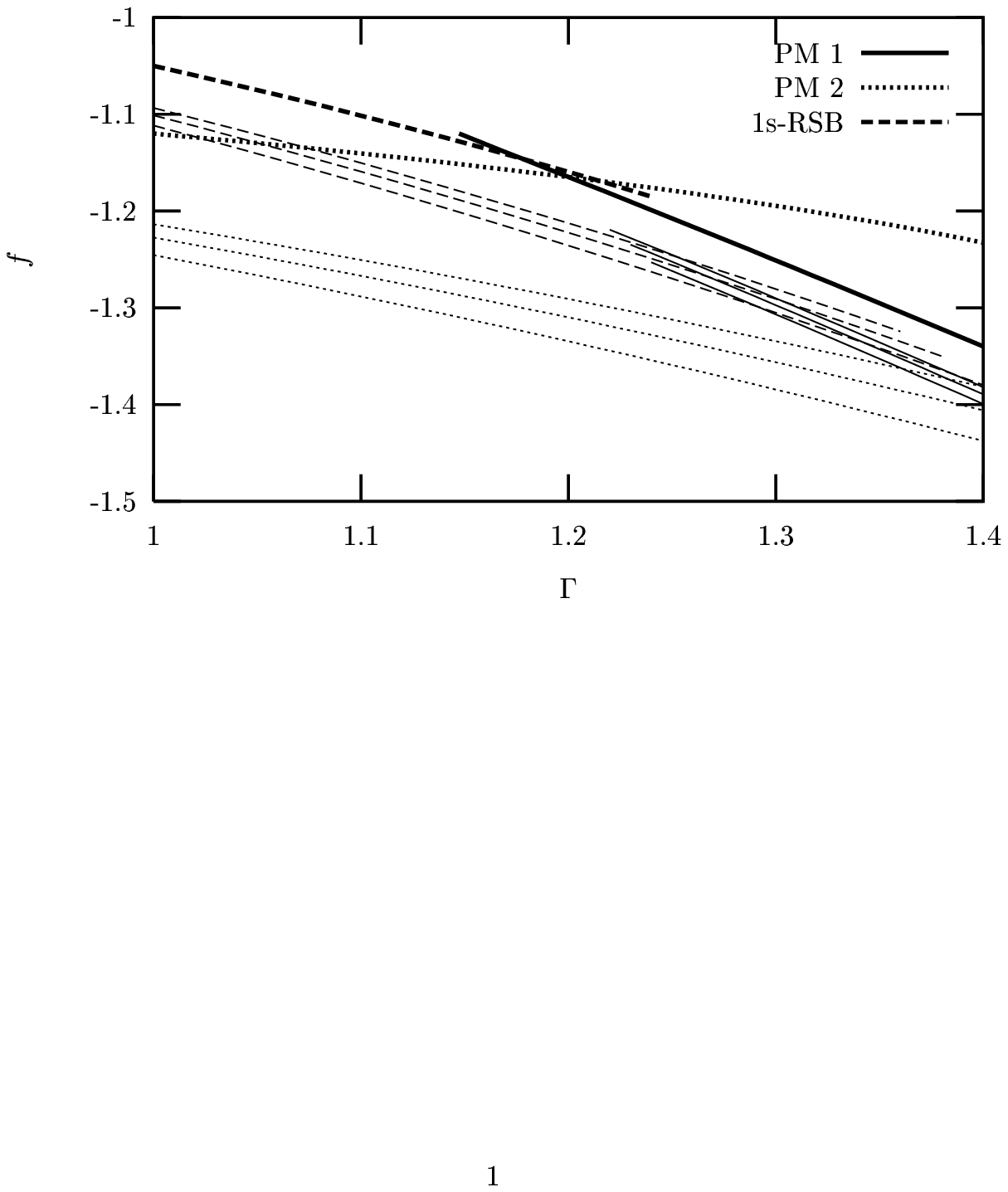}} 
\setcaptionwidth{13cm}
\caption{ \footnotesize Disorder-averaged free-energy density,
 $\overline{f}$, of the quantum $SU(2)$ $p=3$ model at temperature $T =
 0.3$ as a function of the transverse magnetic field $\Gamma$.  The
 coupling to the bath is $\eta=1.0$.  The three phases of the model
 are represented: a physical paramagnet labelled {\sc pm1}, an
 unphysical paramagnet that one discard on physical grounds labeled
 {\sc pm2} and the spin-glass.  The values of $\overline f$ obtained
 for finite $M$ are shown with thin lines [$M=8$ (bottom), $M=9$
 (middle) and $M=10$ (top)]. The result of the extrapolation to $M
 \rightarrow \infty$ is displayed with bold lines.  }
\label{fig:M-dep} 
\end{center} 
\end{figure}

Before discussing the behavior of the model in detail let us
illustrate the dependence on the number of time slices $M$ used in the
{\sc dss} numerical method. In Fig.~\ref{fig:M-dep0} we show the
diagonal order parameter $q_d(\tau)$ as a function of $\tau$ at
constant temperature and transverse magnetic field for the $p=3$
model. Six values of $M$ are shown, $M=8,\cdots,13$. In
Fig.~\ref{fig:M-dep} we display the free-energy density of the
different phases of the $p=3$ model with $\eta=1$.  The four curves
correspond to three values of $M$, $M=8,9,10$, and the result of the
extrapolation to $M\to\infty$. In both cases the variation of the
curves with $M$ is indeed very smooth.
 
Even if the extrapolating method of Goldschmidt and Lai~\cite{Gold} is 
simple to implement and very efficient, it is also limited when the 
temperature decreases. The extrapolation is less clear at lower 
temperatures.  
 
\subsection{Static phases} 
 
Let us first discuss the effect of the environment on the static phase 
diagram of the $p=2$ ({\sc sk}) and $p\geq 3$ quantum $SU(2)$ spin models.  
The spin-glass disorder-averaged free-energy 
densities are obtained using the one-step {\sc rsb} {\it Ansatz} 
discussed in Sect.~\ref{sec:model}.  
We discuss the limits of validity of this 
{\it Ansatz} below. 
 
As in other disordered quantum spin
models~\cite{Niri,Cugrsa1,Cugrsa2,Grempel} two paramagnetic solutions
coexist. As in the spherical $p$-spin model coupled to a bath the one 
labeled {\sc pm2} in Fig.~\ref{fig:M-dep} 
can be discarded since its entropy becomes negative at
sufficiently low temperatures.  Thus, we do not discuss it further in
this paper.
 
The critical line $(T_s,\Gamma_s)$ separating the paramagnetic ({\sc pm}) 
and spin-glass ({\sc sg}) phases is determined by the values of the 
pairs $(T,\Gamma)$ where the physical paramagnetic 
(called {\sc pm1} in Fig.~\ref{fig:M-dep}) and spin-glass free-energy 
densities cross. 
 
\begin{figure} 
\begin{center} 
\resizebox{10cm}{!} 
{\includegraphics*[1.5in,4.7in][7in,7.7in]{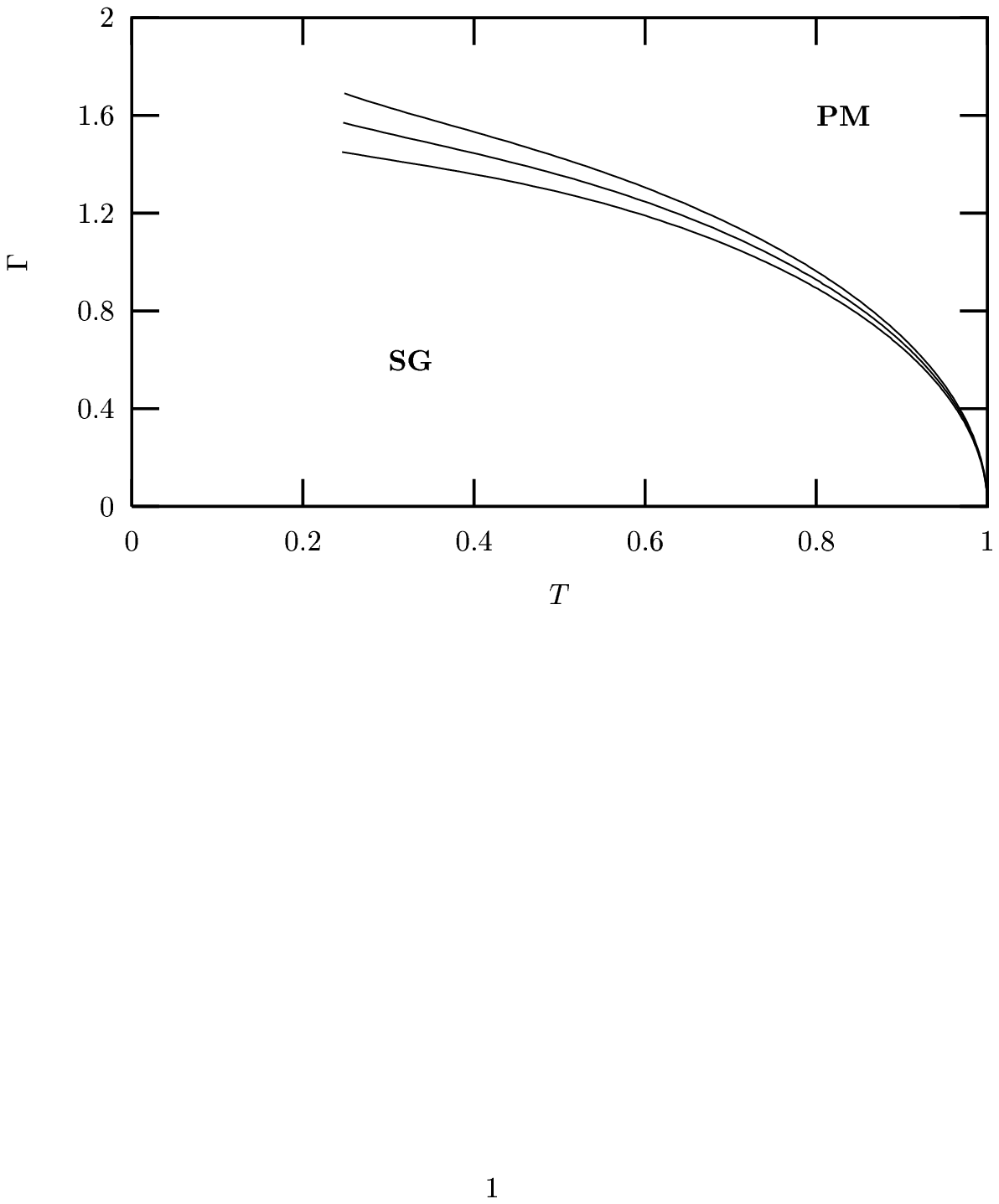}} 
\resizebox{10cm}{!} 
{\includegraphics*[1.5in,4.7in][7in,7.7in]{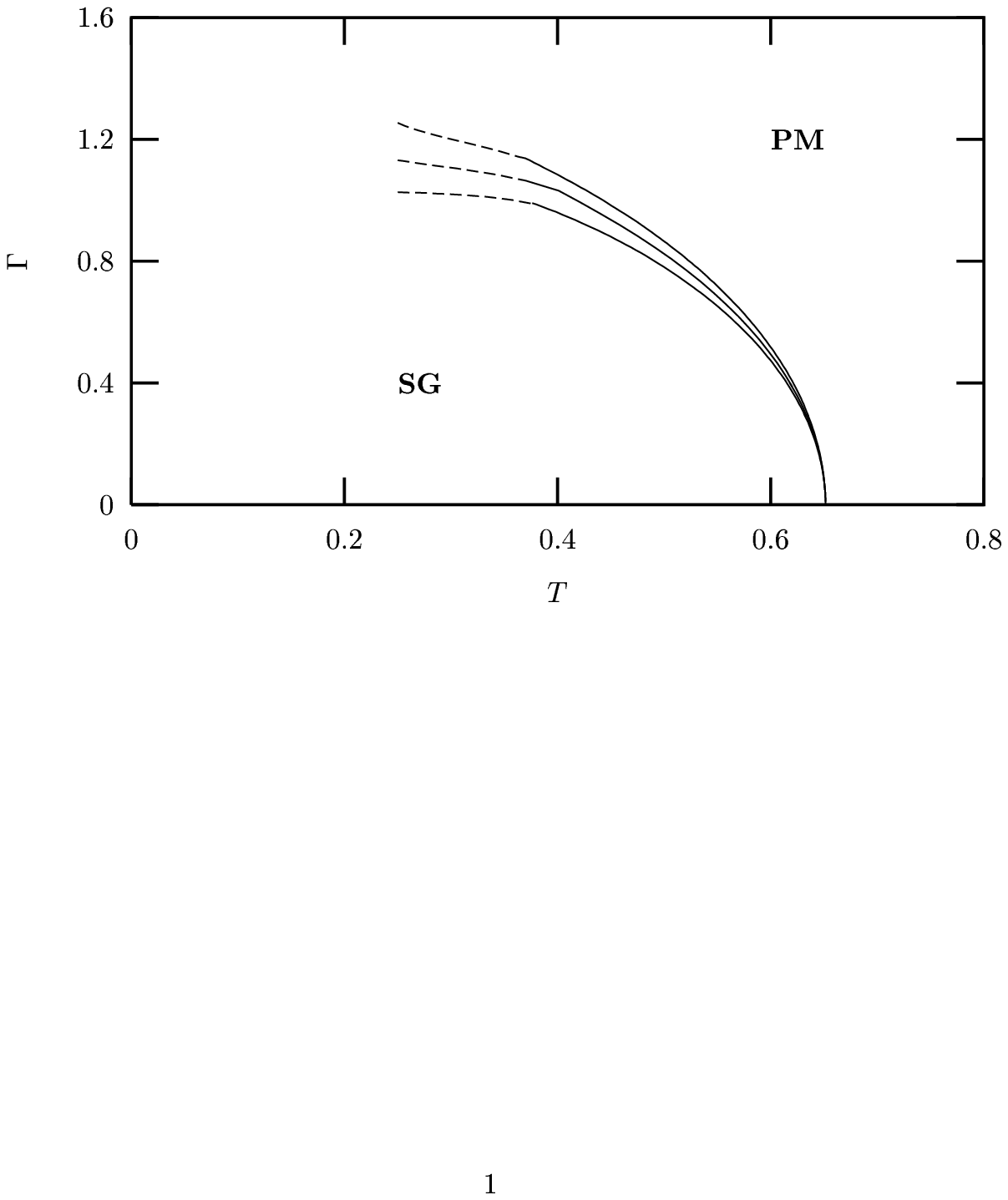}} 
\setcaptionwidth{13cm}
\caption{\footnotesize Upper panel: Static phase diagram for the $p=2$
model as obtained using the {\sc dss} technique for finite number of
time slices $M$ and extrapolating the data to $M \rightarrow \infty$.
The three lines correspond to $\eta=0,0.5,1$, from bottom to top.
Lower panel: Static phase diagram for the $p=3$ model obtained using
the same numerical method.  The continuous line (dashed line)
indicates a second order (first order) phase transition. The critical
lines continue below the lowest value of $T$ for which we trust the
algorithm, $T\approx 0.25$, to reach a quantum critical point at
$T=0$.}
\label{fig1} 
\end{center} 
\end{figure}

We show in the upper and lower panels of Fig.~\ref{fig1} 
the static critical line in the $(T,\Gamma)$ plane separating 
a high $T$, high $\Gamma$ paramagnetic phase ({\sc pm}) from a 
low $T$, low $\Gamma$ spin-glass ({\sc sg}) for the $p=2$ and $p=3$ 
models, respectively. The three curves in each figure correspond to 
($\eta=0$) and two non zero couplings ($\eta = 0.5, 1$) from bottom to 
top. 
 
For both models, the classical transition temperature, $T_s^{class}$, 
corresponding to $\Gamma_s\to 0$, remains unchanged by the
coupling to the quantum heat reservoir.  This value is $T_s=J$ for 
when $p=2$~\cite{Mepavi} and it coincides with the
one given by Gross and M\'ezard, $T_s\approx 0.67$, for the classical
problem with $p=3$~\cite{Gross}.
 
For the three values of $\eta$, the static critical transverse field,
$\Gamma_s(T)$, is a decreasing function of $T$, which is consistent
with the fact that quantum fluctuations tend to destroy the glassy
phase.  We also see from the figures that the coupling to a quantum
thermal bath favors the formation of the glassy phase: the coupling to
the environment effectively reduces the strength of the quantum
fluctuations that tend to destroy it.  For any value of the
temperature that satisfies $T<T_s^{class}$ the extent of the
spin-glass phase is larger for stronger couplings to the
bath. Moreover, we observe that the effect of the bath is stronger for
lower temperatures.
 
\begin{figure} 
\begin{center} 
\resizebox{11cm}{!}{\includegraphics*[1.5in,4.7in][7in,7.7in] 
{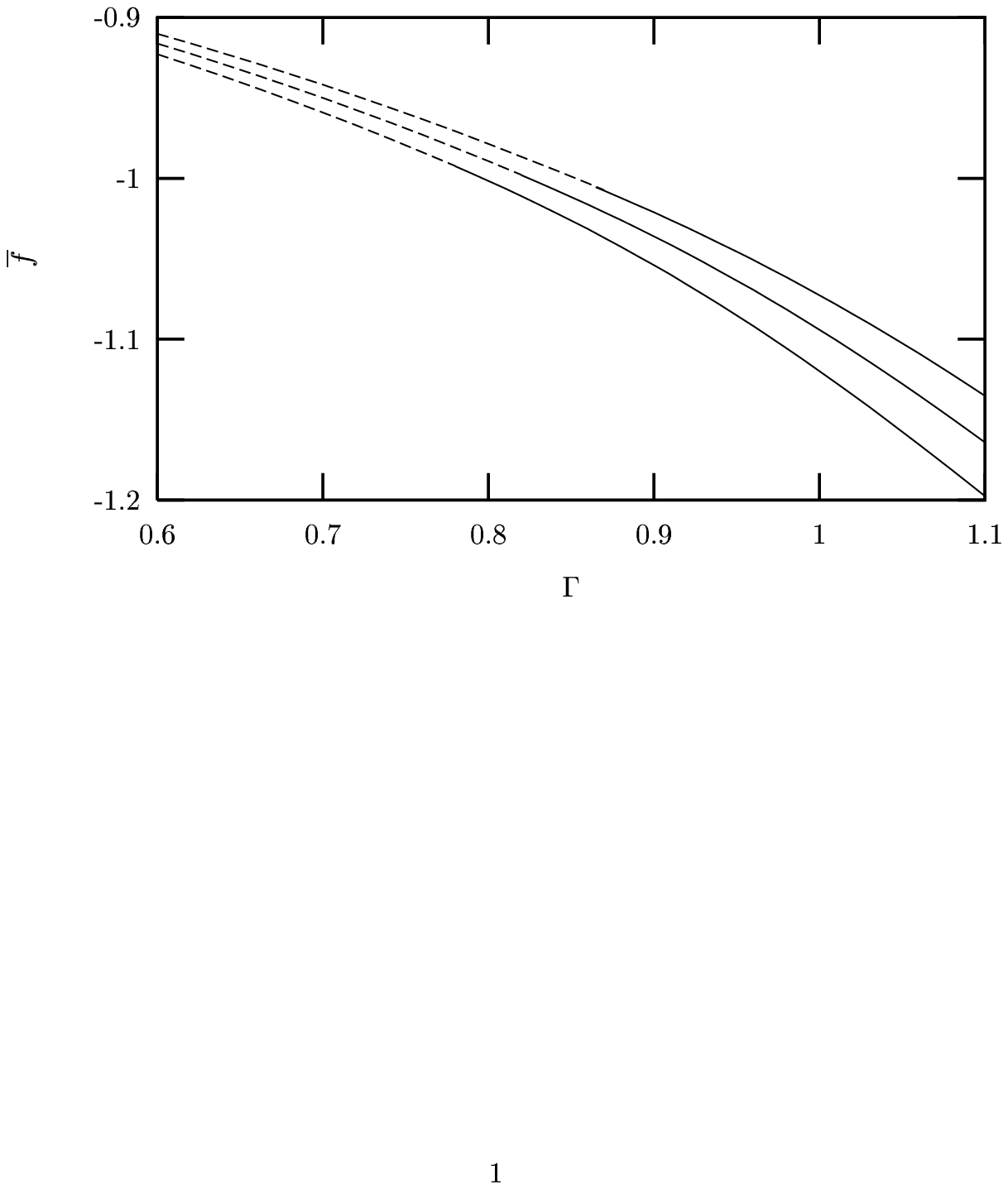}} 
\resizebox{11cm}{!}{\includegraphics*[1.5in,4.7in][7in,7.7in] 
{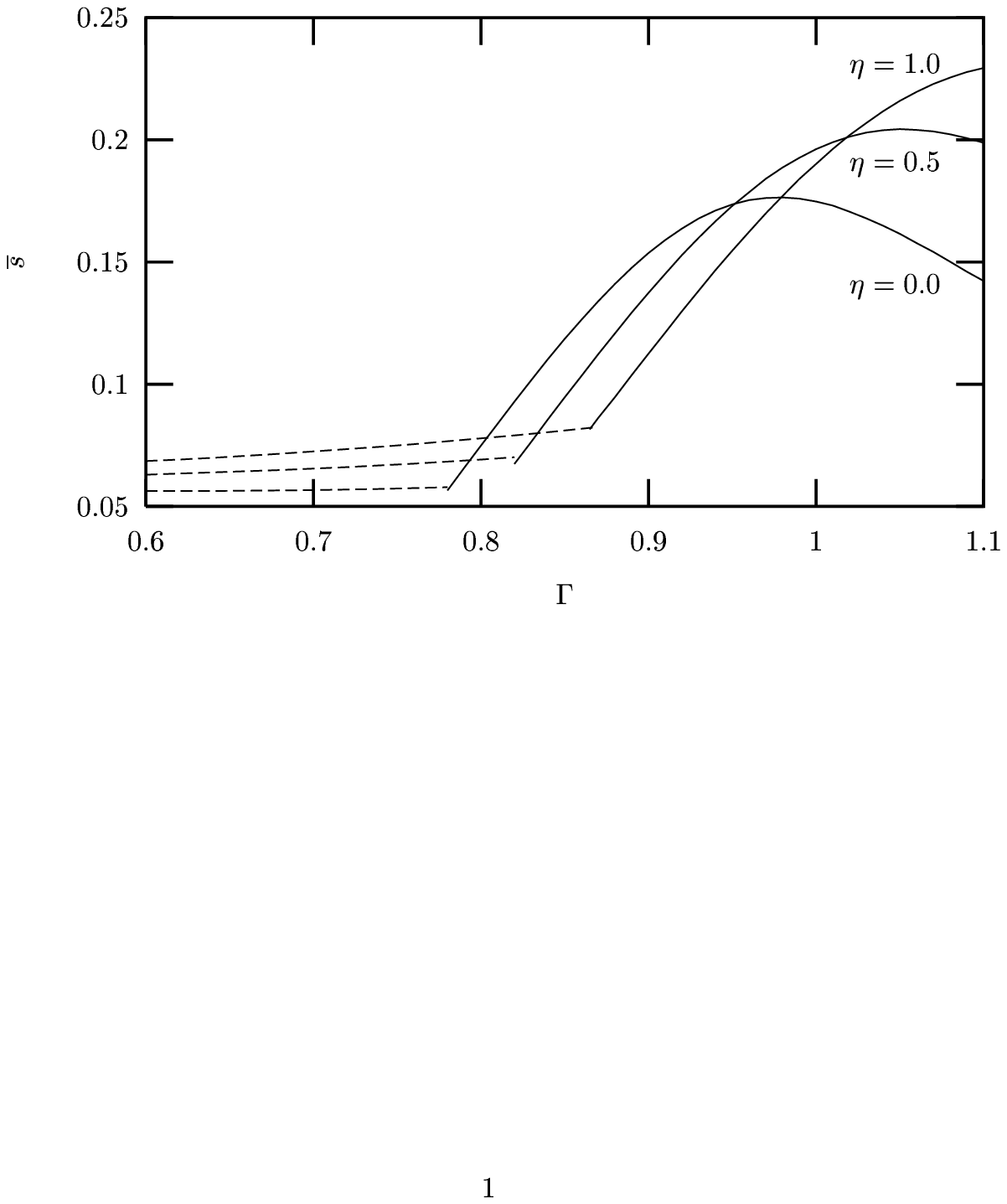}} 
\resizebox{10.5cm}{!}{\includegraphics*[1.5in,4.7in][7in,7.7in] 
{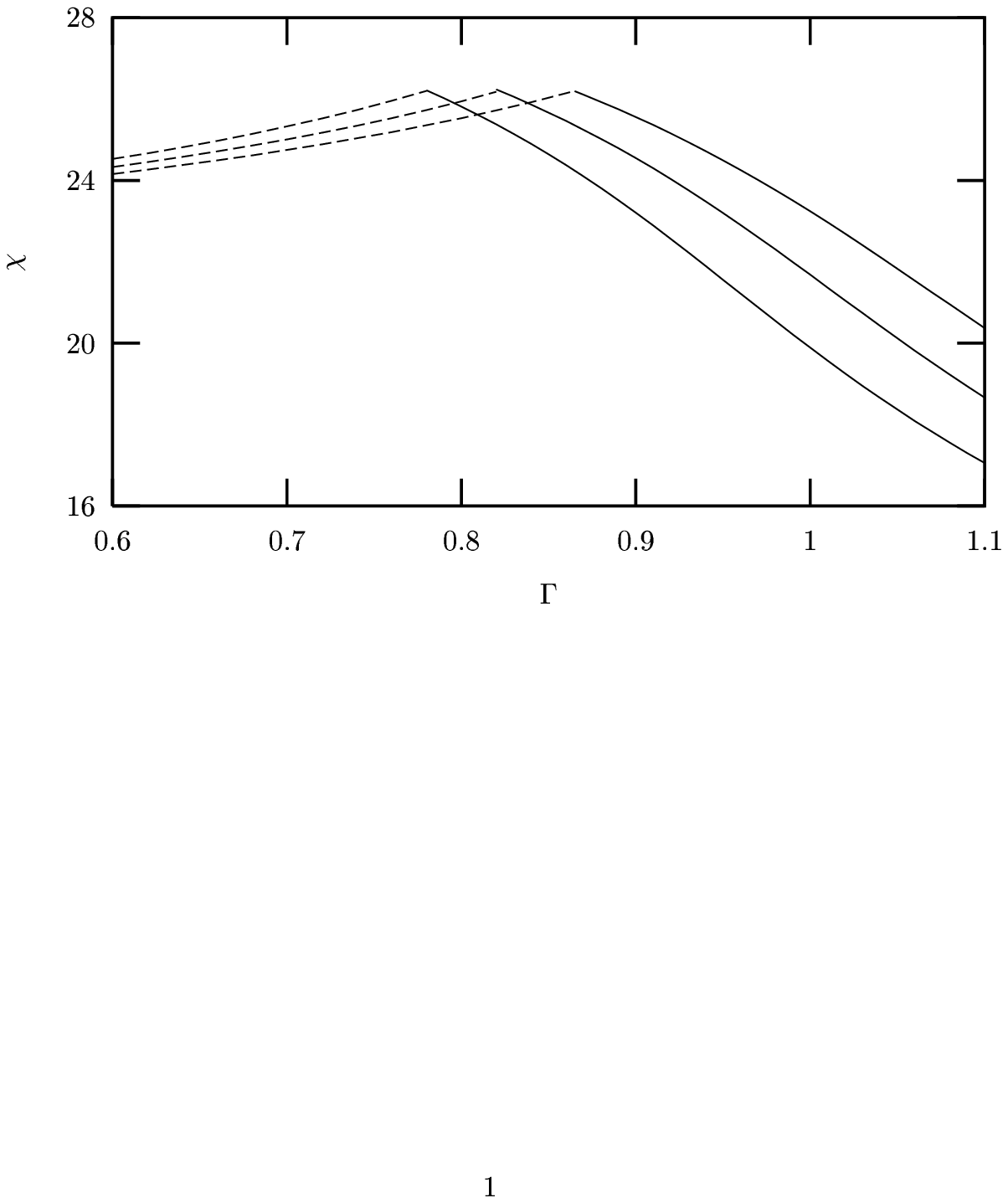}} 
\end{center} 
\setcaptionwidth{13cm}
\caption{ \footnotesize Free energy density, $\overline f$, entropy,
$\overline s$, and susceptibility, $\chi$, as functions of the
transverse field, $\Gamma$, for the $p=3$ model at $T=0.5>T_s^*$ for
three values of the coupling to the bath,
$\eta=0,0.5,1$. The continuous (dashed) line corresponds to the
paramagnetic (glassy) phase. 
The entropy and susceptibility 
are continuous at the transition indicating a second order phase
transition}
\label{fvscampop3con} 
\end{figure}

\begin{figure} 
\begin{center} 
\resizebox{11cm}{!}{\includegraphics*[1.5in,4.7in][7in,7.7in] 
{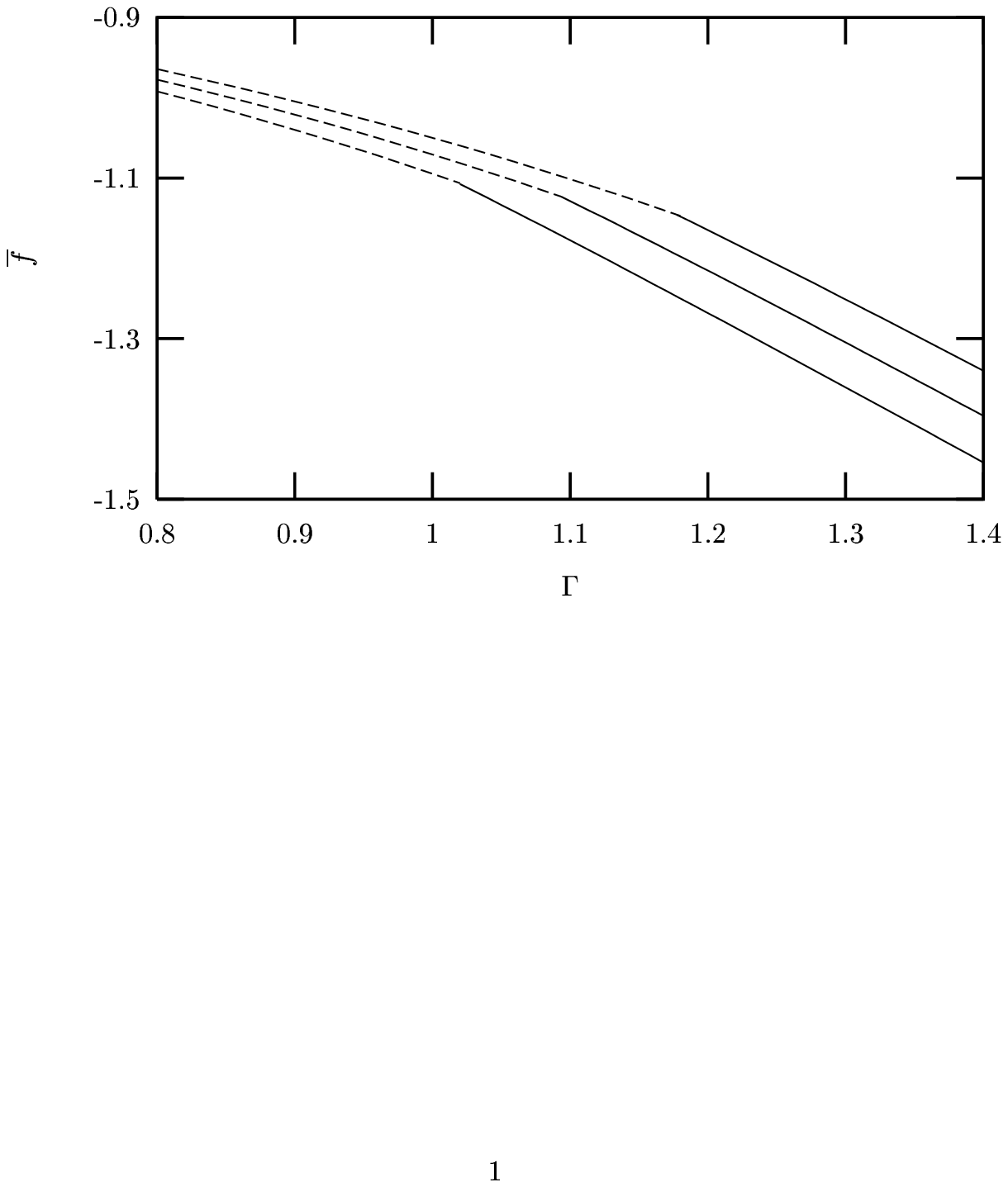}} 
\resizebox{11cm}{!}{\includegraphics*[1.5in,4.7in][7in,7.7in] 
{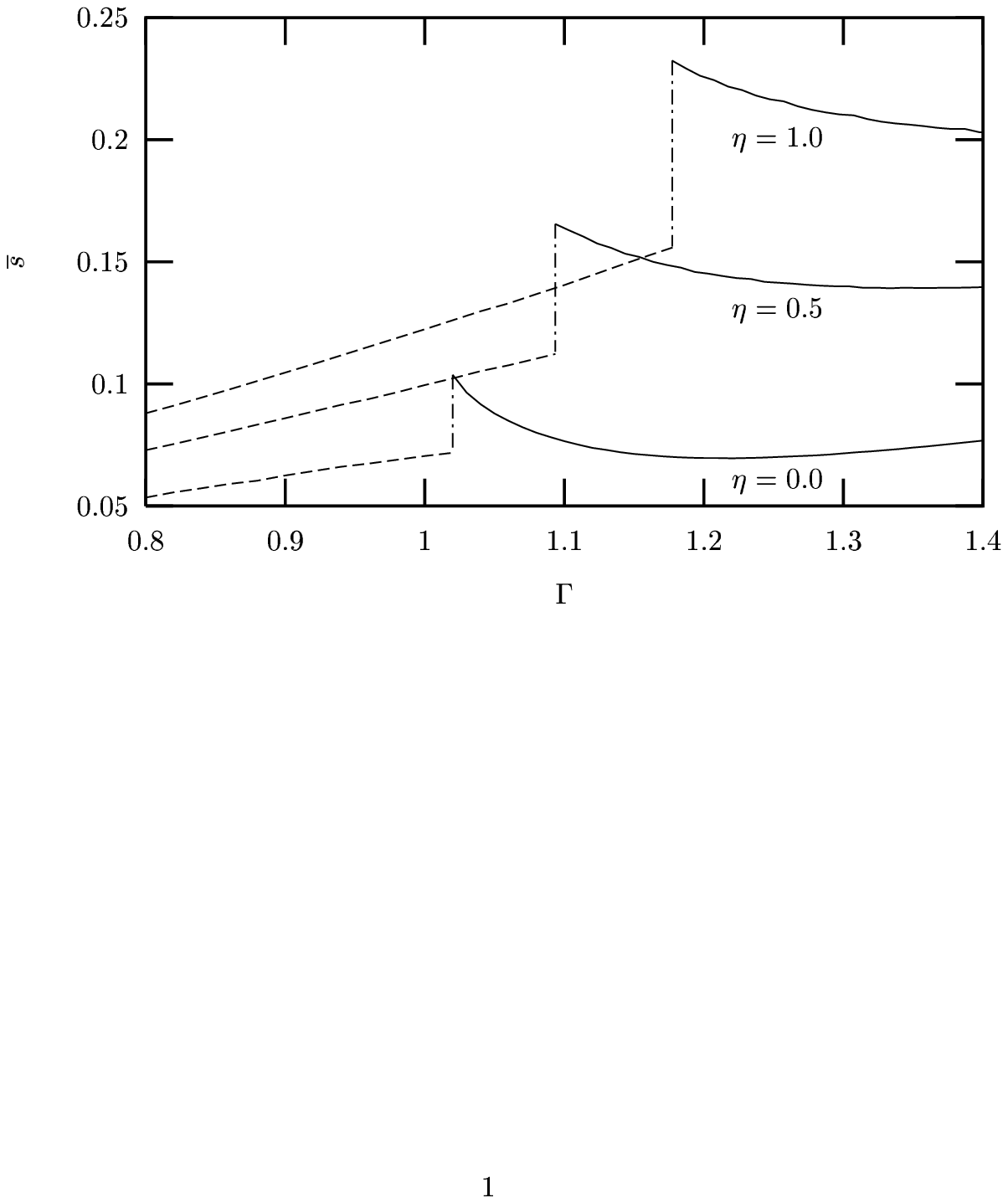}} 
\resizebox{10.5cm}{!}{\includegraphics*[1.5in,4.7in][7in,7.7in] 
{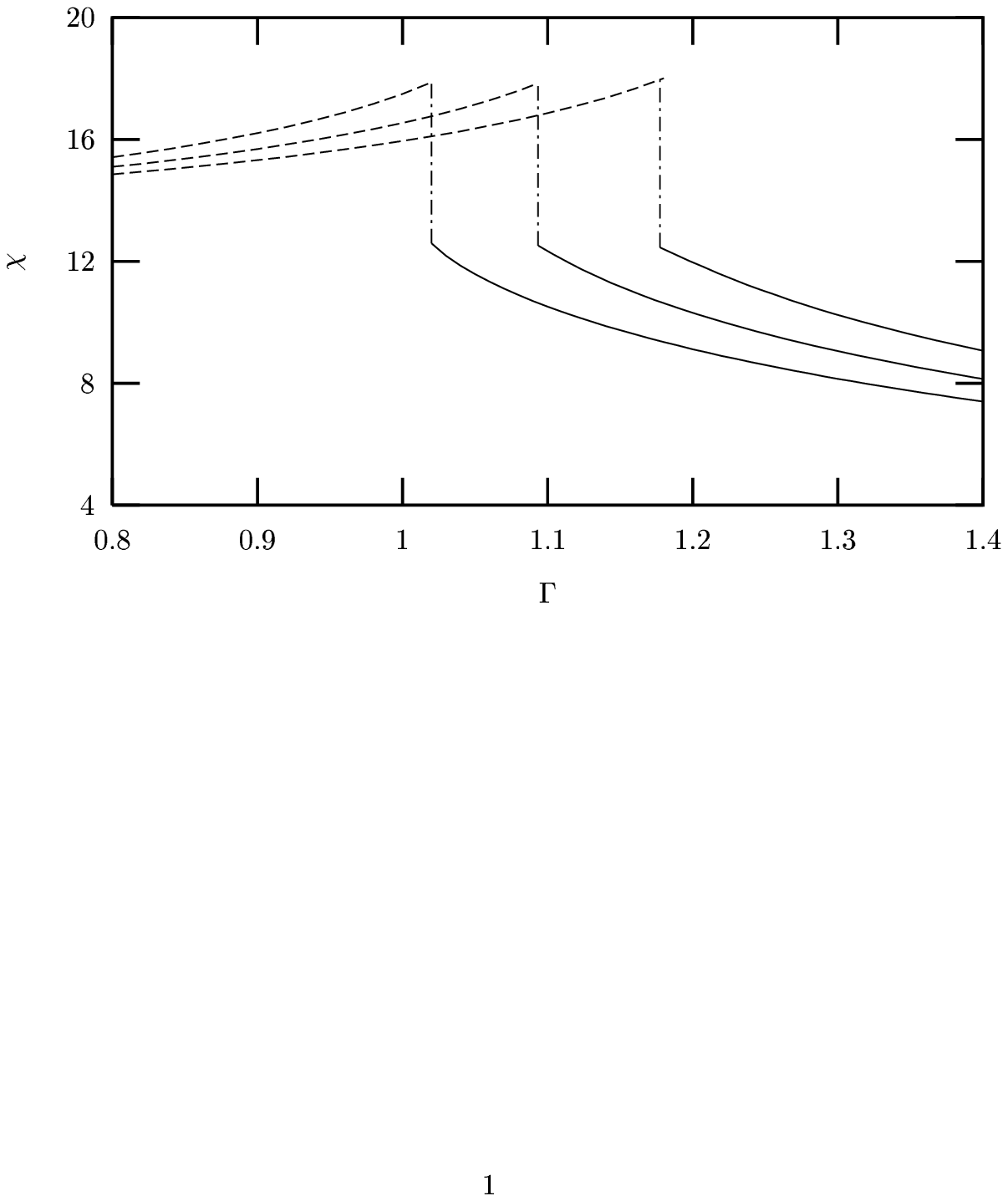}} 
\end{center} 
\setcaptionwidth{13cm}
\caption{\footnotesize Free energy, $\overline f$, entropy, $\overline
s$, and susceptibility, $\chi$, as function of the transverse field,
$\Gamma$, for the $p=3$ model at $T=0.3 < T_s^*$ for three 
values of the coupling to the bath, $\eta=0,0.5,1$. The continuous
(dashed) line corresponds to the paramagnetic (glassy) phase. The
entropy and susceptibility are discontinuous at the transition
indicating a first order phase transition}
\label{fvscampop3} 
\end{figure}

When $p=2$ the transition is always continuous and second-order
thermodynamically. For $p=3$ instead, as in the spherical
case~\cite{Cugrsa1,Cugrsa2,Cugrlolosa} and the isolated quantum $SU(2)$
model~\cite{Niri}, an interesting change from a second-order to a
first-order transition appears.  We demonstrate these statements by
displaying in Figs.~\ref{fvscampop3con} and \ref{fvscampop3} the
behavior of the free-energy density, entropy and susceptibility of the
$p=3$ $SU(2)$ spin model as a function of the transverse field for
$T=0.5>T_s^*$ and $T=0.3<T_s^*$.
 
At sufficiently high temperatures, $T\geq T_s^*$, one finds a 
spin-glass solution for increasing transverse fields until the 
break-point $m$ reaches the value $m=1$. The values 
$(T, \Gamma)$ where $m=1$ coincides with the ones obtained by analyzing 
the crossing of the free-energy densities of the paramagnetic 
and spin-glass solutions.  Thus, for the chosen temperature $T\geq T_s^*$ 
this is the critical transverse field. Even if the Edwards-Anderson 
parameter, $q_{ea}$, and the diagonal element, $q_d(\tau)$, are 
non-zero at this point in parameter space, one can check, as shown in 
Fig.~\ref{fvscampop3con}, that the entropy and susceptibility do not 
show a jump. Thus, for $T\geq T_s^*$ the transition is discontinuous 
[due to the jump in $q_{ea}$ and $q_d(\tau)$] but of second order 
thermodynamically. 
 
The situation is different at lower temperatures. In
Fig.~\ref{fvscampop3} we show the free energy, entropy and
susceptibility of the $p=3$ $SU(2)$ model for $T=0.3<T_s^*$.  In this
case, the point in which the free-energy of the paramagnetic and
spin-glass solution cross corresponds to $m<1$ and as shown in the
figure this leads to a discontinuity of the entropy and
susceptibility. In this case, the transition is discontinuous and
first-order thermodynamically.
 
\begin{figure} 
\begin{center} 
\resizebox{10.5cm}{!} 
{\includegraphics*[1.5in,4.7in][7in,7.7in]{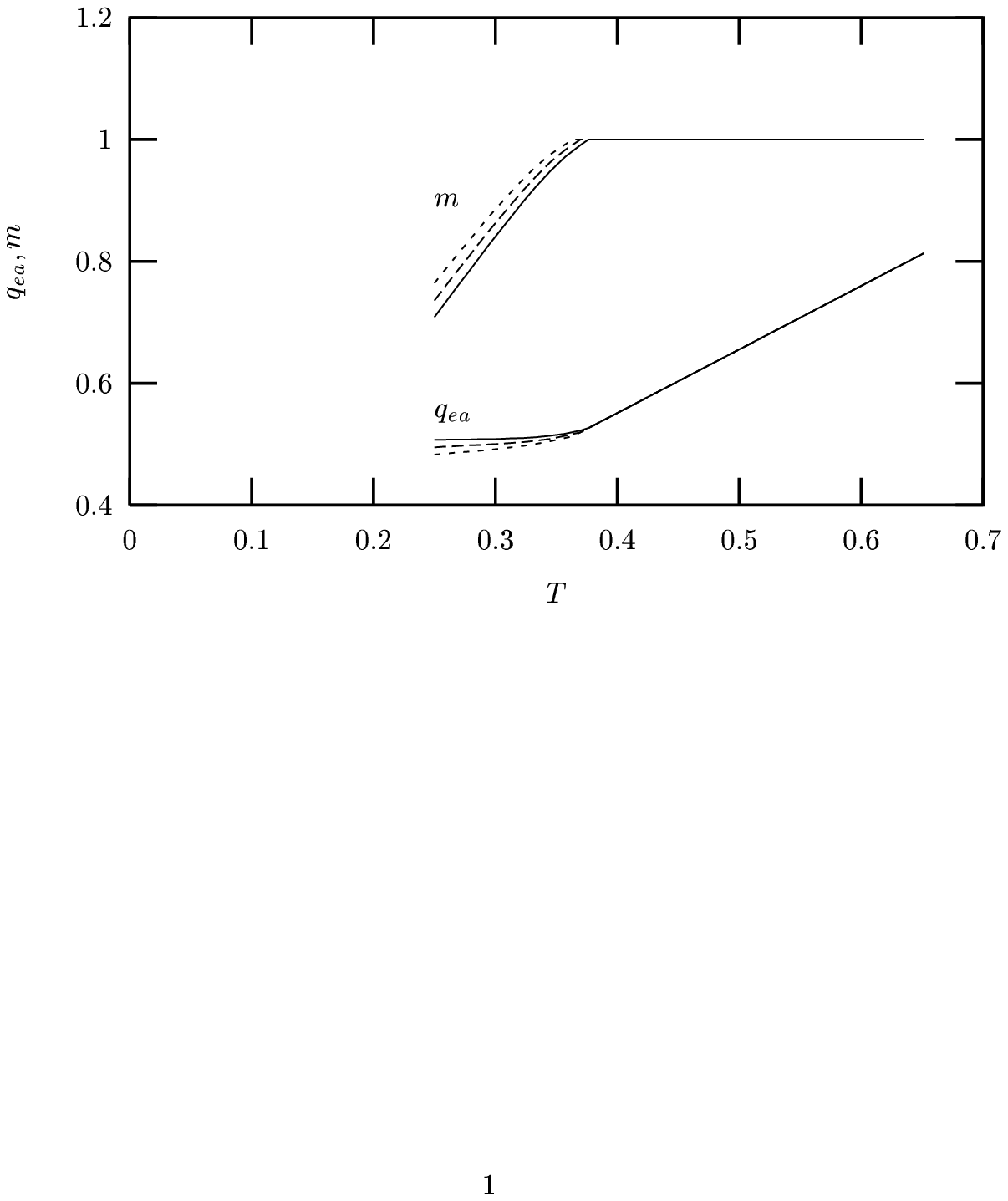}} 
\setcaptionwidth{13cm}
\caption{ \footnotesize Dependence of $m$ and $q_{ea}$ on the
critical temperature $T_s(\Gamma,\eta)$ for the $p=3$ model. Three
values of $\eta$ are considered, $\eta=0,0.5,1$.  For increasing
values of $\eta$ the interval in which $m=1$ increases and, hence, the
region where a thermodynamic first order transition occurs decreases.}
 \label{fig2} 
\end{center} 
\end{figure}

In Fig.~\ref{fig2} we show the dependence of $q_{ea}$ and $m$ on
the critical temperature $T_s$ for three values of the coupling to the
bath. The model is again the $p=3$ quantum $SU(2)$ spin-glass.  As
already mentioned we observe that for all temperatures $q_{ ea} $ is
different from zero, leading to a discontinuous phase transition.  $m$
equals one for $T_s\geq T_s^*$ but $m<1$ for $T_s\leq T^*_s$.  The
figure also shows that $T_s^*$ decreases with increasing coupling to
the bath $\eta$.  Again, this result is reminiscent of what found in
the spherical case~\cite{Cugrlolosa}.
 
\subsection{Stability of the one-step static solution} 
 
In order to study the stability of the one-step solution we evaluated
the replicon eigenvalue $\lambda_R$ on the values of the order
parameters and $m$ obtained from the static solution, and we searched
for the parameters $(T,\Gamma)$ such that $\lambda_R$ vanishes.  In
the classical limit this yields Gardner's classical critical
temperature that takes a rather low value, $T_G(\Gamma=0)\approx
0.25$~\cite{Gardner}. Since we expect to find a decreasing value of
the instability temperature with the strength of the transverse field,
we need to control the numerical algorithm for $T< 0.25$. Even if this
might seem, at first sight, impossible, we managed to obtain sensible
results keeping reachable values of $M$, $M\leq 13$, since the small
values of the transverse field compensate the large value of $\beta$
in the condition $\beta \Gamma/M \ll 1$.
 
First, we analyzed the $p=2$ case that corresponds to the {\sc sk} model 
in a tranverse field. In the absence of the environment we found that the 
one-step {\sc rsb} solution is not stable in the full spin-glass phase 
supporting the idea that the solution to the statics of 
this model needs a full {\sc rsb} scheme, just as in its classical 
limit and in contrast to recent claims in the literature~\cite{Kim}.

\begin{figure} 
\begin{center} 
\resizebox{10.5cm}{!} 
{\includegraphics*[1.5in,4.7in][7in,7.7in]{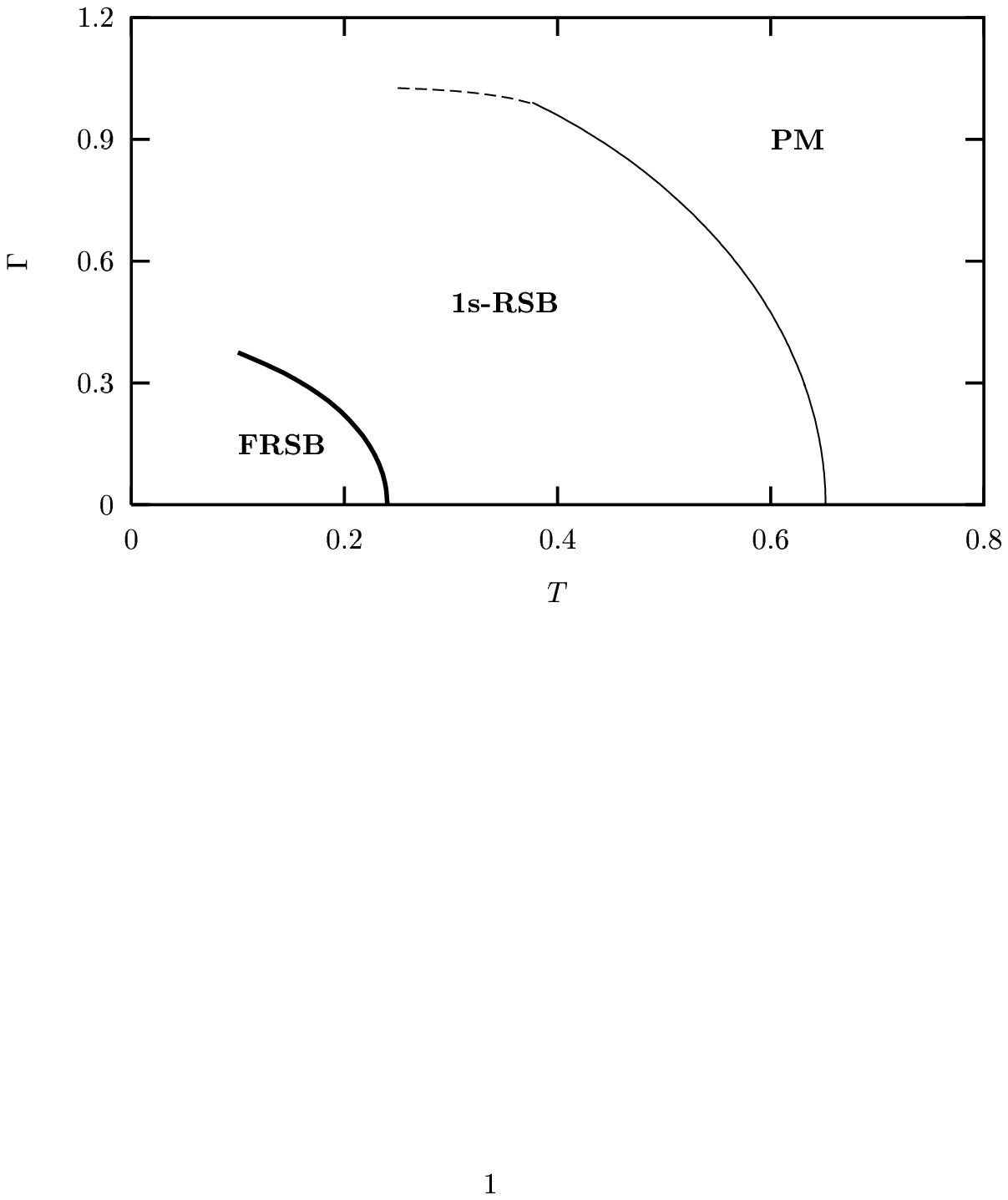}} 
\setcaptionwidth{13cm}
\caption{\footnotesize Comparison between the static critical line
 $(T_s,\Gamma_s)$ and Gardner's instability line $(T_G,\Gamma_G)$ for
 the $p=3$ model with $\eta=0$.}
 \label{fig:Gardner} 
\end{center} 
\end{figure}

In Fig.~\ref{fig:Gardner} we compare the static critical line
$(T_s,\Gamma_s)$ as found from the one-step {\sc rsb} {\it Ansatz},
with Gardner's line of instability for the $p=3$ model.  We see that
the region where the one-step {\sc rsb} static {\it Ansatz} is not
stable is quite small. Since we only trust the extrapolation from low
values of $M$ to $M\to\infty$ above temperatures of the order of
$T\approx 0.1$, we do not explicitly extrapolate the instability line
to lower temperatures. Nevertheless, the existing data suggest that in
the zero temperature limit the static critical transverse field,
$\Gamma_s$, and Gardner's critical field, $\Gamma_G$, do not coincide
$\Gamma_s(T_s=0)> \Gamma_G(T_G=0)$.

\subsection{The dynamic transition} 
 
\begin{figure} 
\begin{center} 
\resizebox{11cm}{!}{\includegraphics*[1.5in,4.7in][7in,7.7in]{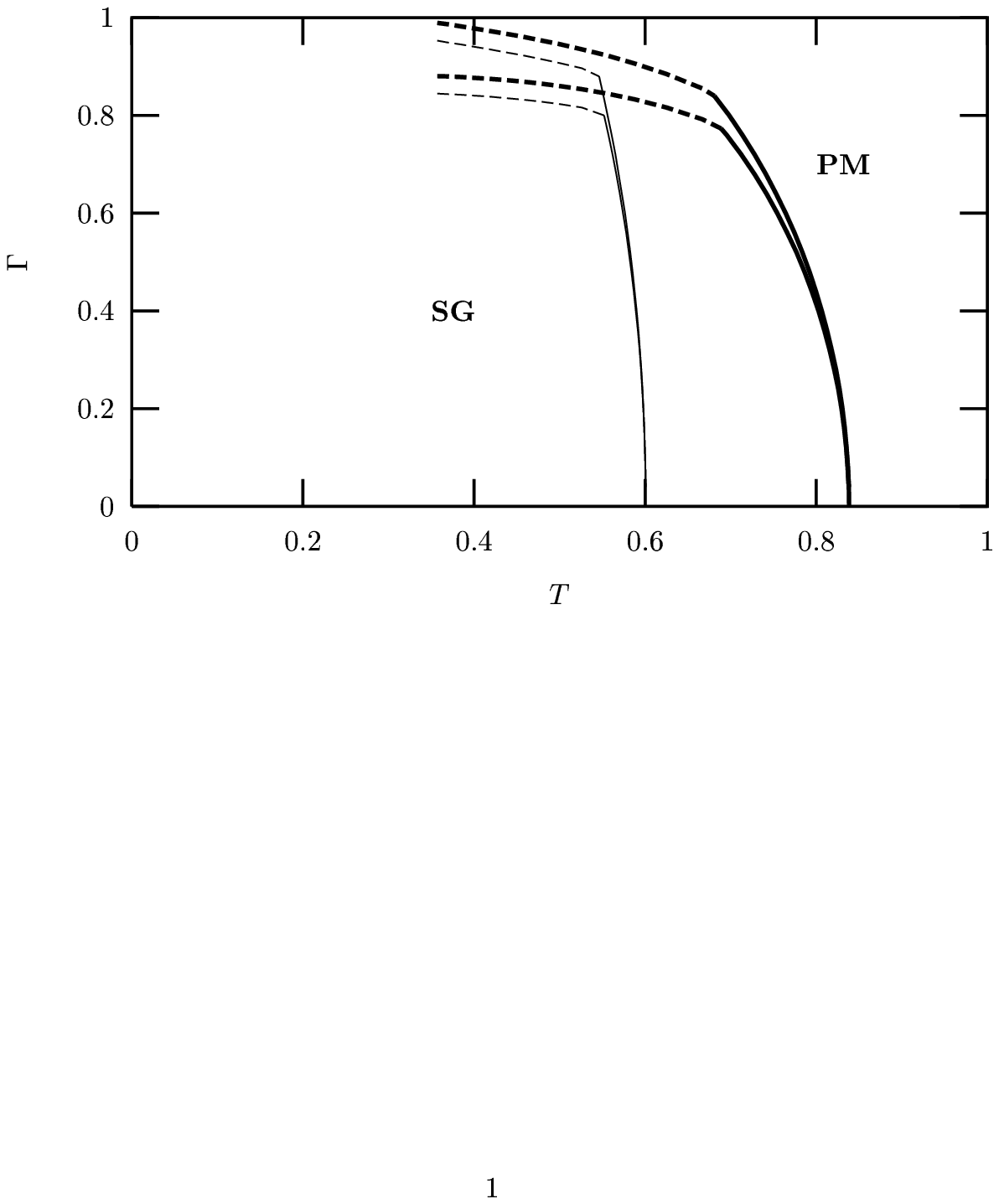}} 
\setcaptionwidth{13cm}
\caption{ \footnotesize Comparison between the static and marginal
critical lines for the $p=10$ model. The solid lines represent second
order transitions and the dashed lines first order ones. The set of
curves with $T_s^{ class}\approx 0.6$ (thin lines) are the static
transition lines for $\eta=0$ (below) and $\eta=1$ (above); the set of
curves with $T_s^{\rm class}\approx 0.82$ (bold lines) correspond to
the dynamic transition for $\eta=0$ (below) and $\eta=1$ (above).}
\label{fig:p10} 
\end{center} 
\end{figure}

As already explained in Sect.~\ref{sec:model} the value of $m$ found
by setting the replicon eigenvalue to zero leads to different equations
that encode some information about the non-equilibrium relaxation
dynamics of the system. Using this prescription we obtained, for the
$p\geq 3$ models a different critical line that lies above the static
transition. This result is similar to those found in a series of other
classical~\cite{pspins} and quantum~\cite{Niri,Cugrsa1,Cugrsa2}
problems. In Fig.~\ref{fig:p10} we compare the static and marginal
critical lines for the $p=10$ quantum $SU(2)$ model. We chose a larger
value of $p$ to make the difference between the two lines easier to
visualize. The glassy static region is smaller than the glassy region
determined by the marginality condition. When approaching the glassy
phase from any direction in parameter space, the dynamic transition,
associated to the line of marginal stability, occurs before the static
one.  As on the critical static line, the curve determined with the
marginal stability criterion is made of two pieces, on one of them the
transition is of second-order (indicated with a solid line on
Fig.~\ref{fig:p10}) and on the other the transition is of first-order
(indicated with a dashed line on the same figure). The first-order
nature of the dynamic transition is displayed by, for instance, a jump
in the asymptotic value of the averaged internal energy. The marginal
tri\-cri\-tical point occurs at higher temperature than the static one.
 
The external noise also has a strong effect on the dynamic critical line. 
The stronger the coupling to the environment (larger value of $\eta$), the 
larger the spin-glass region in the phase diagram. This is also shown in 
Fig.~\ref{fig:p10} where a couple of curves, corresponding to 
$\eta=0$ and $\eta=0.5$ are drawn (see the caption in the figure for more 
details). 
 
Finally, let us mention that there is an empirical relation between the 
value of the parameter $m$ as found from the marginality condition and 
how the fluctuation-dissipation theorem is modified in the real-time 
non-equilibrium relaxation of the quantum model~\cite{Culo,Cugrlolosa}. 
Using this relation and interpreting then 
the parameter $m/T$ as an effective temperature~\cite{Cukupe} we find 
that the modification of the fluctuation-dissipation
theorem, and hence $T_{ eff}$, depend 
on the strength of the coupling to the bath. 
 
\section{Conclusions} 
\label{sec:conc} 
 
In this paper we studied the effect of an Ohmic quantum bath on the
statics and dynamics of quantum disordered $SU(2)$ spin models of
mean-field type. We found that the coupling to the enviromemnt favors
the appearance of the spin-glass phase reducing the strength of the
quantum fluctuations that tend to destabilize it.  As in the case of
the spherical model~\cite{Cugrlolosa,Chandra} the phase transition is
always second order for $p=2$. For $p \ge 3$ there exists a tricritial
temperature $T^\star$ below which quantum fluctuations drive the
transition first order. $T^\star$ decreases with the strength of the
coupling to the bath.  For $p \ge 3$ a dynamic precedes the
equilibrium phase transition. The coupling to the bath also stabilizes
the dynamic glassy phase.
 
It would be interesting to check if the same tendency to ordering
appears in macroscopic spin models in finite dimensions. One could
attempt to study this problem in the context of frustrated spin
magnets or the much studied, numerically and analytically, quantum
$SU(2)$ spin chain with and without disorder.
 
This problem is of interest for possible implementations of 
quantum computers where the interaction of the system with its 
environment needs to be controlled. The effect of an environment on the 
properties of Griffith phases has also been the focus of a hot 
debate~\cite{Griffiths}. We expect to report on these problems in the 
future. 
 
\vspace{2cm} 
 
\noindent 
\underline{Acknowledegments} 
We acknowledge financial support from the an Ecos-Sud travel grant, 
the ACI project "Optimisation algorithms and quantum disordered 
systems". LFC is research associate at ICTP Trieste and 
acknowledges financial support from the J. S. Guggenheim Foundation.
G. S. L. is supported by 
EPSRC grants GR/N19359 and GR/R70309. This research was supported in part 
by the National Science Foundation under Grant No. PHYS99-07949. 
We thank F. Ritort for very 
useful discussions.


\begin{thebibliography}{99} 
 
\bibitem{review-Leggett} A. J. Leggett, S. Chakravarty, A. T. Dorsey, 
M. P. A. Fisher, A. Garg and W. Zwerger, Rev. Mod. Phys. 
{\bf 59}, 1 (1987); {\bf 67}, 725 (1995). 
 
\bibitem{Weiss} U. Weiss in {\it Series Modern Condensed 
Matter Physics} (World Scientific, Sigapore, 1993) Vol. 2. 
 
\bibitem{aeppli} W. Wu, B. Ellmann, T. F. Rosenbaum, G. Aeppli, and 
D. H. Reich,  Phys. Rev. Lett. {\bf 67} 2076 (1991); 
W. Wu, D. Bitko, T. F. Rosenbaum, and G. Aeppli,  Phys. Rev. Lett. {\bf 71} 
1919 (1993). 
 
\bibitem{lohn} R. Vollmer, T. Pietrus, H. v. L\"{o}hneysen, 
R. Chau, and M. B. Maple, Phys. Rev. B {\bf 61}, 1218 (2000). 
 
\bibitem{tabata} Y. Tabata, D. R. Grempel, M. Ocio, T. Taniguchi, and 
Y. Miyako, Phys. Rev. Lett. {\bf 86}, 524 (2001). 
 
\bibitem{ladieu} F. Ladieu, J. Le Cochec, P. Pari, P. 
Trouslard, and P. Ailloud, 
 Phys. Rev. Lett. {\bf 90}, 205501 (2003). 
S. Ludwig and D. D. Osheroff, 
    Phys. Rev. Lett. {\bf 91}, 105501 (2003). 
S. Rogge, D. Natelson, and D. D. Osheroff, 
    Phys. Rev. Lett. {\bf 76}, 3136 (1996). 
 
\bibitem{mean-field} 
D. R. Grempel, M.J. Rozenberg, Phys. Rev. Lett. {\bf 79}, 389  (1998). 
M.J. Rozenberg and D. R. Grempel Phys. Rev. Lett. {\bf 81}, 2550 (1998). 
 
\bibitem{review-quantum-sg} 
H. Rieger and A. P. Young, {\it Quantum spin glasses} 
(Springer-Verlag, Berlin, 1996), cond-mat/9607005. 
R. Bhatt, {\it Quantum spin glasses} in ``Spin-glasses and 
random fields'', A. P. Young ed. (World Scientific, Singapore, 1997). 
 
 
\bibitem{Culo} 
L. F. Cugliandolo, and G. Lozano, 
Phys. Rev. Lett. {\bf 80}, 4979 (1998), 
Phys. Rev. B {\bf 59}, 915 (1999). 
 
\bibitem{Cugrlolosa} 
L. F. Cugliandolo, D. R. Grempel, G. Lozano, H. Lozza, and C. A. da Silva 
Santos, Phys. Rev. B {\bf 66}, 014444 (2002). 
 
 
\bibitem{Chandra} M. Rokhni, and P. Chandra, cond-mat/0301166. 
 
\bibitem{Bipa} G. Biroli and O. Parcollet, 
Phys. Rev. B {\bf 65}, 094414 (2002). 
 
\bibitem{Pottier} N. Pottier, and A. Mauger, Physica A {\bf 282} 
77 (2000). 
 
 
\bibitem{Kondo} 
D. R. Grempel and M. J. Rozenberg, Phys. Rev. B {\bf 60}, 4702 (1999). 
 
\bibitem{Niri} T. M. Nieuewenhuizen and F. Ritort, 
Physica A {\bf 250}, 89 (1998). 
 
\bibitem{Cugrsa1} L. F. Cugliandolo, D. R. Grempel, and C. A. da Silva 
Santos, Phys. Rev. Lett. {\bf 85}, 2589 (2000). 
 
\bibitem{Cugrsa2} L. F. Cugliandolo, D. R. Grempel, and C. A. da Silva 
Santos, Phys. Rev. B {\bf 64}, 014403 (2001). 
 
\bibitem{Bicu} G. Biroli, and L. F. Cugliandolo, Phys. Rev. B {\bf 64}, 
014206 (2001). 
 
\bibitem{Gold} Y. Y. Goldschmidt and P-Y Lai, 
Phys. Rev. Lett. {\bf 64}, 2467 (1990). 
P-Y Lai and Y. Y. Goldschmidt, Europhys. Lett. {\bf 13}, 289 (1990). 
 
\bibitem{Feve} R. P. Feynman and Vernon, Jr. 
Ann. Phys. (NY), {\bf 24}, 118 (1963); 
R. P. Feynman, {\it Statistical mechanics}. Addison Wesley, 1972. 
\bibitem{Leggett} 
A. Caldeira and A. Leggett, Phys. Rev. Lett. {\bf 46}, 211 
A. Caldeira and A. Leggett, Ann Phys. {\bf 149}, 374 (1983). 
 
 
\bibitem{Grempel} 
D. R. Grempel and M. J. Rozenberg, Phys. Rev. B {\bf 60}, 4702 (1999). 
 
\bibitem{Mepavi} M. M\'ezard, G. Parisi and M. A. Virasoro, 
{\it Spin glass theory and beyond} (World Scientific, Singapore, 1987). 
 
\bibitem{Bray-Moore} A. J. Bray and M. A. Moore, J. Phys. C {\bf 13},
L655 (1980).
 
 
\bibitem{pspins}
T. R. Kirkpatrick and D. Thirumalai, Phys. Rev. B 
{\bf 36}, 5388 (1987). T. R. Kirkpatrick and P. G. Wolynes, 
Phys. Rev. B {\bf 36}, 8552 (1987).   
L. F. Cugliandolo and J. Kurchan, Phys. Rev. Lett. 
{\bf 71}, 173 (1993). 
 
\bibitem{quantum-marg}
T. Giamarchi and P. Le Doussal,  Phys. Rev. B {\bf 53} 15206 (1996).
  A. Georges, O. Parcollet and S. Sachdev, 
Phys. Rev. Lett., {\bf 85}, 840 (2000), Phys. Rev. B {\bf 63}, 134406 (2001).
G. Schehr, T. Giamarchi and P. Le Doussal, cond-mat/0212300.

 
\bibitem{replicon} A. J. Bray and M. A. Moore, 
J. Phys C {\bf 12}, L441 (1979). 
 
\bibitem{AT} J. R. L.  de Almeida and D. J. Thouless, J. Phys. 
A {\bf 11}, 983 (1978). 
 
\bibitem{Gardner} E. Gardner, Nucl. Phys. B {\bf 257} [FS14], 747 (1985). 
 
\bibitem{ussc} K. D. Usadel and B. Schmitz, Solid State 
Comm. {\bf 64}, 975 (1987). 
 
\bibitem{alri} J. V. Alvarez and F. Ritort, J. Phys. A: 
Math. Gen. {\bf 29}, 7355 (1996). 
 
\bibitem{suzuki} M. Suzuki, Prog. Theor. Phys., 56,1454 (1976) 
 
 
\bibitem{Gross} D. J. Gross and M. M\'ezard, Nucl. 
Phys. B {\bf 240} [FS12], 431 (1984). 
 
\bibitem{Kim} D-H Kim and J-J Kim, 
Phys. Rev. B {\bf 66}, 054432 (2002). 
 
\bibitem{Cukupe} L. F. Cugliandolo, J. Kurchan and L. Peliti, 
Phys. Rev. E {\bf 55}, 3898 (1997). 
 
\bibitem{Griffiths} A. H. Castro-Neto and B. A. Jones 
Phys. Rev. B {\bf 62}, 14975 (2000). 
A. J. Millis, D. Morr and J. Schmalian, 
Phys. Rev. B {\bf 66}, 174433 (2002).
 
\end{thebibliography}
\end{document}